\documentclass{llncs}

\usepackage{cite}
\usepackage{graphicx}
\usepackage{amsfonts}
\usepackage{amssymb}
\usepackage{amsmath}
\usepackage{calc}
\usepackage{textcomp}
\usepackage{ifthen}
\usepackage{pdfcomment}
\usepackage[dvipsnames]{xcolor}
\usepackage{tikz}
\usepackage{booktabs}
\usepackage{tabularx}
\usepackage{paralist}
\usepackage{algorithm}
\usepackage{algorithmicx}
\usepackage[noend]{algpseudocode}
\usepackage{threeparttable}
\usepackage{todonotes}
\usepackage{subcaption}
\usepackage{caption}
\usepackage{cleveref}
\usepackage{csquotes}

\usepackage{array}

\newcolumntype{L}[1]{>{\raggedright\let\newline\\\arraybackslash\hspace{0pt}}m{#1}}
\newcolumntype{R}{@{\extracolsep{7pt}}l@{\extracolsep{7pt}}}

\def\BibTeX{{\rm B\kern-.05em{\sc i\kern-.025em b}\kern-.08em
    T\kern-.1667em\lower.7ex\hbox{E}\kern-.125emX}}

\usepackage{etoolbox}
\makeatletter
\providecommand{\subtitle}[1]{\apptocmd{\@title}{\par {\large #1 \par}}{}{}
}
\makeatother
\begin{document}

\title{Quantifying the Re-identification Risk\\of Event Logs for Process Mining}
\subtitle{Empiricial Evaluation Paper}

\author{Saskia Nu\~nez von Voigt\inst{1} \and Stephan A.
Fahrenkrog-Petersen\inst{2} \and\\ Dominik Janssen\inst{3} \and Agnes
Koschmider\inst{3} \and Florian Tschorsch\inst{1} \and\\ Felix Mannhardt\inst{4,5} \and Olaf Landsiedel\inst{3}  \and Matthias Weidlich\inst{2}}

\institute{
    Technische Universit\"at Berlin, Germany \\
    \email{\{saskia.nunezvonvoigt,florian.tschorsch\}@tu-berlin.de}
    \and
    Humboldt-Universit\"at zu Berlin, Germany\\ \email{\{stephan.fahrenkrog-petersen,matthias.weidlich\}@hu-berlin.de}
    \and
    Kiel University, Germany\\\email{doj|ak|ol@informatik.uni-kiel.de}
\and
	SINTEF Digital, Trondheim, Norway\\
     \email{felix.mannhardt@sintef.no}\\
     \and
	NTNU Norwegian University of Science and Technology, Trondheim, Norway\\
    }

\maketitle

\begin{abstract}
  Event logs recorded during the execution of business processes
  constitute a valuable source of information. Applying process mining
  techniques to them, event logs may reveal the actual process
  execution and enable reasoning on quantitative or qualitative process
  properties. However, event logs often contain
  sensitive information that could be related to individual process
  stakeholders through background information and cross-correlation.
  We therefore argue that, when publishing event logs, the risk of such
  re-identification attacks must be considered. In this paper, we show how to
  quantify the re-identification risk with measures for the individual
  uniqueness in event logs. We also report on a large-scale study that explored
  the individual uniqueness in a collection of publicly available event logs.
  Our results suggest that potentially up to all of the cases in an event log may be
  re-identified, which highlights the importance of privacy-preserving
  techniques in process mining.
\end{abstract}

\setcounter{footnote}{0}

\section{Introduction}
\label{intro}

Process mining uses data recorded in the form of event logs by information
systems to, for example, reveal the actual execution of business
processes~\cite{DBLP:books/sp/Aalst16}.
Since most activities in modern organization are supported by technology,
each process execution produces a digital footprint indicating the occurrence
and timing of activities.
Consequentially, event logs may contain sensitive information and are vulnerable to adversarial attacks.
Unfortunately, there is no general method on how to safely remove personal and sensitive references.
Since the existence of privacy threats are generally known,
the willingness to publish event logs is low.
Publicly available event logs, however, are necessary
to evaluate process mining models~\cite{Mannhardt2019,bauerELPaaSEventLog2019,icpm/Fahrenkrog-Petersen19}
and therefore discussions are needed on how to safely publish event logs.
Against this background, we argue that it is crucial to understand the
risk of data re-identification in event logs and process mining.
With this insight, we can balance how much information of an event log can be
shared and how much should be anonymized to preserve privacy.
While many examples confirm the general risk of data re-identification~\cite{7821808,7796899,Rocher2019},
the re-identification risk of event logs has not received much attention yet.

The intention of this paper is to raise awareness to the re-identification risk of event logs
and therefore provide measures to quantify this risk.
To this end, we provide an approach to express the \emph{uniqueness} of data,
which is derived from models that are commonly adopted by
process mining techniques. Each event recorded in an event log consists of
specific data types, such as the activity name of the respective process
step, the timestamp of its execution, and event attributes
that capture the context and the parameters of the activity.
Additionally, sequences of events that relate to the same case of a
process, also known as traces, come with data attributes, so-called
case attributes that contain general information about the case.
To extract sensitive information, an adversary uses background knowledge
to link a target's attributes with the case/event attributes in the event log,
e.g., by cross-correlating publicly-available sources.
The higher the uniqueness of an event log, the higher an adversary's chances
to identify the target.
Our approach therefore explores the number of cases that are uniquely identifiable by the
set of case attributes or the set of event attributes.
We use this information to derive a measure of uniqueness for an event log,
which serves as a basis for estimating how likely a case can be re-identified.

To demonstrate the importance of uniqueness considerations for event logs,
we conducted a large-scale study with 12 publicly available event logs
from the 4TU.Centre for Research Data
repository.\footnote{\url{https://data.4tu.nl/repository/collection:event\_logs\_real}}
We categorized the records and assessed the uniqueness where cases refer to a natural person.
Our results for these logs suggest that an adversary
can potentially re-identify up to all of the cases, depending on prior
knowledge. We show that an adversary needs only a few attributes of a trace
to successfully mount such an attack.

\noindent The contributions of this paper can be summarized as follows:
\begin{compactitem}
	\item We present an approach to quantify the privacy risk associated to event
	logs. In this way, we support the identification of information that
	should be suppressed when publishing an event log, thereby fostering the
	responsible use of logs and paving the way for novel use cases based
	on event log analysis.
\item By reporting the results of a large-scale evaluation study, we
  highlight the need to develop privacy-preserving
  techniques for event logs with high utility for process analytics. Our
  notions of individual uniqueness may serve as a catalyst for such efforts,
  since they make the inherent privacy risks explicit.
\end{compactitem}
This paper is structured as follows.
\Cref{privacy} illustrates privacy threats in process mining.
\Cref{approach} presents our approach for quantifying the re-identification 
risk.
We analyze publicly available event logs and discuss the results in 
\Cref{results}.
We review related work in \Cref{related}, before \Cref{conclude}
concludes this paper.

 \section{Privacy Threats in Process Mining}
\label{privacy}
Process mining uses event logs to discover and analyze business processes.
Event logs capture the execution of activities as events.
A finite sequence of such events forms a trace, representing a single process 
instance (aka case). For example, the treatment of patients in an emergency 
room includes a number of events, such as blood sampling and analysis,
which together follow a certain structure as determined by the process.
Accordingly, the events related to an individual patient form a case.
In addition, case attributes provide general information about a case,
e.g., place of birth of a patient.
Each event consists of various data types, such as the name of the respective \emph{activity},
the \emph{timestamp} of the execution, and \emph{event attributes}.
Event attributes are event-specific and may be changing over time,
e.g., a temperature or the department performing a treatment.
The key difference between case attributes and event attributes is
that case attributes do \emph{not} change their value
for a case during the observed period of time.
We show a synthetic event log example capturing an emergency room process in 
\Cref{tab:event_log_example}.

\begin{table}[tb]
    \caption{Event log example}
    \label{tab:event_log_example}
    \begin{tabularx}{\textwidth}{l l l X X}
        \toprule
            case id & activity & timestamp & case attributes & event attributes \\
        \midrule
1000 & registration\hspace{1ex} & 03/03/19 23:40:32\hspace{1ex} & 
            \{age: 26, sex: m\}& \{arrival: check-in\}  \\
            1000 & triage & 03/04/19 00:27:12 & \{age: 26, sex: m\}& \{status: 
            uncritical\} \\
            1000 & liquid & 03/04/19 00:47:44& \{age: 26, sex: m\}& \{liquid: 
            NaCl\} \\
            ... & ... & ... & ... & ... \\
            1001 & registration & 03/04/19 00:01:24 &\{age: 78, sex: f\} & 
            \{arrival: ambulance\} \\
1001 & antibiotics &03/04/19 00:09:06 &\{age: 78, sex: f\}  & 
            \{drug: penicillin\} \\
            ... & ... & ... & ... & ... \\
        \bottomrule
    \end{tabularx}
\end{table}

Considering the structure of an event log, several privacy threats are 
identified. Linking a case to an individual can reveal sensitive information, 
e.g., in an emergency room process, certain events can indicate that a patient 
is in a certain condition. In general, case attributes can contain various 
kinds of sensitive data,
revealing racial or ethnic origin, political opinions, religious or philosophical beliefs,
as well as financial or health information. Likewise, an event log can reveal information about the productivity~\cite{pika2017mining} or the work schedule of hospital staff. Such kind of staff surveillance is a critical privacy threat. Clearly, it is essential to include privacy considerations in process mining projects. We assume that an adversary's goal is to identify an individual in an event log linking external information. Depending on the type of background information, 
different adversary models are possible. We assume a targeted 
re-identification, i.e., an adversary has information about specific 
individuals, which includes a subset of the attribute values. Based thereon, 
the adversary aims to reveal sensitive information, e.g., a diagnosis.
Here, we assume that an adversary knows that an individual is present in the 
event log. In this paper, we consider the uniqueness measure to quantify the 
re-identification risk of sensitive information, thereby providing a basis for 
managing privacy considerations.

\section{Re-identifications of Event Logs}
\label{approach}

To apply our uniqueness measure to cases, we summarize all occurring event data to its corresponding case.
This assumption eases handling multiple events belonging to the same case. Since case attributes are invariant over time, they only need to be taken into consideration once, whereas event attributes may be different for every event and therefore their temporal change needs to be considered. \Cref{tab:data_preparation} provides a respective example.
Each row in this table belongs to one case.
The case attributes \enquote{sex} and \enquote{age} are listed in separate columns.
The columns \enquote{activity}, \enquote{timestamp}, and \enquote{arrival channel}
contain an ordered list of the respective attributes.
For example, the case id 11 has only two events and therefore two activities.
The second activity \enquote{antibiotics} on March~04, 2019 has no \enquote{arrival channel} (i.e., it is \enquote{none}).
\begin{table}[tb]
    	\caption{Preparation of event log}
    	\label{tab:data_preparation}
    	\setlength{\tabcolsep}{2.2pt}
    \begin{tabularx}{\textwidth}{c c c l l l }

        \toprule
            case id & sex & age & \multicolumn{1}{>{\centering\arraybackslash}c}{activity} & \multicolumn{1}{>{\centering\arraybackslash}c}{timestamp}  & \multicolumn{1}{>{\centering\arraybackslash}c}{arrival channel} \\
        \midrule
            10 & male & 26 & [reg., liquid, $\dots$]  & [3/3/19, 3/4/19, $\dots$] & [check-in, none, $\dots$]  \\
            11 &female& 78 & [reg., antibiotics] & [3/4/19, 3/4/19]  & [ambulance, none]\\
            12 &female& 26 & [reg., liquid, $\dots$] & [3/5/19, 3/7/19, $\dots$]  & [check-in, none, $\dots$]\\
            $\cdots$ & $\cdots$ & $\cdots$& \multicolumn{1}{>{\centering\arraybackslash}c}{$\cdots$}   & \multicolumn{1}{>{\centering\arraybackslash}c}{$\cdots$}  & \multicolumn{1}{>{\centering\arraybackslash}c}{$\cdots$} \\
            \bottomrule
    \end{tabularx}
\end{table}

The uniqueness of an event log serves as a basis for estimating how likely a case can be re-identified. We investigate a number of so-called projections that can be considered as a data minimization technique, effectively reducing the potential risks of re-identification in an event log.
Projections refer to a subset of attributes in the event log. They can easily be adopted to assess the risk in different scenarios. \Cref{tab:projections} summarizes the projections for event logs and their potential usage in process mining.
Projection~A contains the sequence of all executed activities with their timestamps,
while projection~F only contains the case attributes. It has been shown that even sparse projections of event logs hold privacy risks~\cite{icpm/Fahrenkrog-Petersen19}. Therefore, in our evaluation, we will consider the re-identification risk for various projections.

\subsection{Uniqueness based on case attributes}
\label{sec:uniqueness_attributes}
In addition to unique identifiers (UID),
so-called quasi-identifiers are information that can be linked to individuals as well.
A combination of quasi-identifiers may be sufficient to create a UID.
In event logs, the case attributes can be seen as quasi-identifier.
For example, in the event log of the BPI Challenge 2018~\cite{BPI18}, the area of all
parcels and the ID of the local department can be considered as case attributes.
Measuring the uniqueness based on case attributes is a common way to quantify
the re-identification risk~\cite{dankar2012estimating}.
Case uniqueness and thus an individual uniqueness highly increases the risk of re-identification.
A single value of a case attribute does not lead to
identification. The combination with other attributes, however, may lead to a unique case.
In particular, when linking attributes to other sources of information,
it may result in successful re-identification.

\begin{table}[tb]
	\caption{Projections of event logs}
	\label{tab:projections}
\begin{tabularx}{\textwidth}{ R l R}
		\toprule
		projection & data included & exemplary usage in process mining \\
		\midrule
A & activities, timestamps & analysis of bottlenecks \\
		B & activities, event and case attributes & predictive process monitoring\\
		C & activities, event attributes & decision mining\\
		D & activities, case attributes & trace clustering \\
		E & activities & process discovery \\
		F & case attributes & traditional data mining\\
		\bottomrule
	\end{tabularx}
\end{table}

We define uniqueness as the fraction of unique cases in a event log. Let $f_k$
be the frequency of the $k$th combination of case attributes values in a
sample. One case is unique if $f_k=1$, i.e., there is no other case
with the same values of case attributes.
Accordingly, uniqueness for case attributes is defined as
\begin{equation}
    U_\text{case}=\frac{\sum I(f_k=1)}{N},
\end{equation}
where the indicator function $I(f_k=1)$ is $1$, if the $k$th combination is unique,
and $N$ is the total number of cases in the event log.
Referring to our data in \Cref{tab:data_preparation},
the attribute value \enquote{sex:~female} leads to two possible case candidates (id:10 and id:11),
i.e., $f_k=2$, which implies that the combination is not unique.
Taking \enquote{age} as an additional quasi-identifier into account, makes all three listed cases unique,
i.e., $U_\text{case} = 1$. Since often a sample of the event log is published,
we distinguish between sample uniqueness and population uniqueness. The number of unique cases in the sample is
called sample uniqueness. With population uniqueness, we refer to the amount of
unique cases in the complete event log (i.e., population).
Based on the disclosed event log we can measure the sample uniqueness.
The population uniqueness is the number of cases that are unique within the sample
and are also unique in the underlying population from which the data has been sampled.
Usually the event log is a sample from a population and the original event log is
not available.
Therefore, the population uniqueness cannot be measured and must be estimated.

There are several models to estimate the population
uniqueness from a sample. These methods model the population uniqueness
based on extrapolations of the contingency table to fit specific distributions
to frequency counts~\cite{dankar2012estimating}.
We adopt the method of Rocher and Hendrickx~\cite{Rocher2019} to estimate
the population uniqueness.\footnote{Code available at 
\url{https://github.com/computationalprivacy}.}
The authors use Gaussian copulas to model population
uniqueness, approximate the marginals from the sample,
and estimate the likelihood for a sample unique being a population unique.
For this analysis, we assume that the event log is a published sample.
By applying the method, we estimate the population uniqueness of cases
in terms of their case attributes.

\subsection{Uniqueness based on traces}
\label{sec:uniqueness_traces}
Most of the published event logs for process mining do not have many
case attributes, only event attributes.
For example, the Sepsis event log~\cite{Mannhardt.2016} has only one case attribute (\enquote{age}).
However, a case can also be unique based on the events.
We measure the uniqueness using the traces. A trace consists of an ordered set of activities $a_1,a_2,\dots a_n$,
their timestamps $t_1,t_2,\dots t_n$
and $l$ event attributes $e_{11},\dots e_{ln}$.
A tuple $p_j = (a_j,t_j,e_{1j}$,\ldots,$e_{lj})$ represents a
point from the trace
$[(a_1,t_1,e_{11},...,e_{l1})$,$(a_2,t_2,e_{12},...,e_{l2})$,\ldots,$(a_n,t_n,e_{1n},...,e_{ln})]$.
We assume that an adversary's main goal is to re-identify an individual given a
number of points and to reveal other sensitive points.
We argue that an adversary has a certain knowledge and knows some points,
which she is able to link with the event log.
In particular, we assume that an adversary knows that a certain person is contained in the event log.
In other words, we consider the published event log as population.
As our example in \Cref{tab:data_preparation} shows, even without considering the
case attributes, all cases are unique:
Case 11 is uniquely identifiable by its second activity \enquote{antibiotics}.
The Cases 10 and 12 are uniquely identified by combining the activity with the respective timestamp.
An adversary for example might have information about a patient's arrival
(e.g., \enquote{check-in: 3/5/19}).
Given this information as a point from the trace it is sufficient for an adversary to identify the patient
and reveal additional information from the event log.

Accordingly, we express the re-identification risk as the ratio of unique cases.
The uniqueness of a trace can be measured similarly
to location trajectories~\cite{song2014not,de2013unique}.
In location trajectories, points consist only of a location and a timestamp.
In contrast, we have not only two-dimensional but multi-dimensional points
with i.a. an activity, a resource, and a timestamp.
Let $\{c_i\}_{i=1,\dots,N}$ be the event consisting of a set of $N$ traces.
Given a set of $m$ random points, called $M_p$ we compute the number of
traces that include the set of points.
A trace is unique if the set of points $M_p$ is only contained
in a single trace.
The uniqueness of traces given $M_p$ is defined as
\begin{equation}
    U_\text{trace} = \frac{\sum \delta_i}{N},
\end{equation}
where $\delta_i=1$, if a trace is unique
$|\{c_i | M_p \subseteq c_i \} |=1$,
otherwise $\delta_i=0$.  \section{Results}
\label{results}
For our evaluation we used the publicly available event logs from the 4TU.Centre
for Research Data. We classified the event logs into real-life-individuals (R) and
software (S) event logs.
The case identifier of real-life-individuals refers to a natural person,
e.g., the ADL event log~\cite{ActivitiesDaily} includes
activities of daily living activities of individuals.
In event logs referring to software activities, events do not directly refer to a natural person,
but to technical components.
For instance, the BPI Challenge 2013 event log~\cite{BPI13} consists of events from an incident management system.
Some of the software related event logs even consist of a single case,
which makes measuring uniqueness of cases more difficult.
However, if a suitable identifier can be linked to the cases, it will also be
possible to measure the uniqueness for software related event logs.
For example, the incidents in the BPI Challenge 2013 event log are processed
by a natural person.
By using an appropriate transformation, this natural person could serve as a case identifier.

In the following, we apply our methods to estimate the uniqueness of the
real-life-individuals event logs (R) only.
We measure the uniqueness of case attributes for
event logs with more than one case attribute only.
\Cref{tab:categorization} summarizes the results of our classification,
provides some basic metrics on the number of cases and activities,
and indicates the applied uniqueness measures.

\begin{table}[tb]
    \caption{Classification of event logs}
    \label{tab:categorization}
    \centering
    \setlength{\tabcolsep}{7pt}
    \begin{tabularx}{\textwidth}{L{3cm} c r r c c}
    \toprule
     & & & & \multicolumn{2}{c}{uniqueness}\\
    \cmidrule{5-6}
    event log & category & \#cases & \#activities & case attr. & traces  \\
    \midrule
    ADL \cite{ActivitiesDaily} & R & 75 & 34 & no&yes\\
    BPIC 2012 \cite{BPI12}  & R & 13,087 & 24 &yes&yes\\
    BPIC 2015 \cite{BPI15} & R & 1,199 & 398 &yes&yes\\
    BPIC 2017 \cite{BPI17}& R & 31,509 & 26 &yes&yes\\
    BPIC 2018 \cite{BPI18}& R &  43,809 & 14 &yes&yes\\
    CCC 2019 \cite{CCC19} & R & 10,035 & 8 &no&yes\\
    Credit \cite{Credit17} & R & 20 & 29 &no&yes\\
    HB \cite{Hospital17} & R & 100,000 & 18 &no&yes\\
    RlH \cite{HospitelReal11} & R & 1,143 & 624 & no & yes\\
    WABO \cite{WABO14}  & R & 1,434 & 27 & yes & yes \\
    RTFM \cite{Road15} & R & 150,370 & 11 & no & yes \\
    Sepsis \cite{Mannhardt.2016} & R & 1,049 & 16 & no & yes \\
    \midrule
    Apache \cite{Apache}& S & 3 & 74 &-&-\\
    BPIC 2013 \cite{BPI13} & S & 1,487 & 4 &-&-\\
    BPIC 2014 \cite{BPI14}& S & 46,616 & 39 &-&-\\
    BPIC 2016 \cite{BPI16}& S & 25,647 & 600 &-&-\\
    BPIC 2019 \cite{BPI19}& S & 251,734 & 42 &-&-\\
    JUnit \cite{JUnit16} & S & 1 & 182 &-&-\\
    NASA \cite{NASA17} & S & 2,566 & 47 &-&-\\
    SWA \cite{Statechart18} & S & 1 & 106 &-&-\\
\bottomrule
\end{tabularx}
\end{table}

For improved readability and for ethical considerations (see \Cref{discuss} for details),
we will apply our methods and
discuss intermediate results in detail only for the BPI Challenge 2018~\cite{BPI18} and
the Sepsis~\cite{Mannhardt.2016} event logs.
For all other event logs, we provide condensed and pseudonymized results.
Note that the pseudonymized event logs in the following sections
have not the same order as in \Cref{tab:categorization},
but the pseudonymization is consistent across the evaluation.

\subsection{Uniqueness results based on case attributes}

The BPI Challenge 2018 event log is provided by the German company \enquote{data experts}.
It contains events related to application of
payments process of EU's Agricultural Guarantee Fund.
The event log consists of 43,809 cases, each representing a farmer's direct
payments application over a period of three years.
We identified \enquote{payment\_actual}~(PYMT), \enquote{area} (ARA), \enquote{department}~(DPT),
\enquote{number\_parcels}~(\#PCL), \enquote{smallfarmer}~(SF), \enquote{youngfarmer}~(YF),
\enquote{year}~(Y) and \enquote{amount\_applied}~(AMT) as case attributes.
The data contributor generalized the attributes PYMT, \#PCL, and AMT
by grouping the values in 100 bins, where the bins are identified by the 
minimum value~\cite{BPI18}.

To determine the impact of case attributes, we evaluate their uniqueness using 
various combinations.
Specifically, we investigate which combinations of attribute values
make cases more distinct and thus unique. The more extensive an adversary's 
background knowledge is, 
the more likely it is that this individual becomes unique and thus identifiable.
For each combination, we count the number of unique cases.
As expected, the more case attributes are known, the more unique the cases become.
\Cref{tab:bpi18qicombi} (left) shows that when considering PYMT only, there are 40.9\% unique cases.
In combination with \#PCL, uniqueness increases to 69.8\%.
With all case attributes, 84.5\% of the cases are unique in the sample.

\begin{table}[tb]
\centering
\caption{Sample uniqueness and population uniqueness (estimated) based on case attributes (left for BPI Challenge 2018; right for all event logs)}
\label{tab:bpi18qicombi}\label{tab:results_qi}
\begin{minipage}[t]{0.6\textwidth}\centering
    \resizebox{1\textwidth}{!}{
    \setlength{\tabcolsep}{5pt}
    \begin{tabularx}{1.1\textwidth}{X r r}
    \toprule
    combination & sample & population \\
    \midrule
    PYMT  & 0.409 & 0.161 \\
    PYMT, ARA &  0.476 &  0.164\\
    PYMT, DPT &  0.528& 0.419 \\
    PYMT, \#PCL &  0.698& 0.594 \\
    PYMT, ARA, \#PCL &        0.747 & 0.649 \\
    PYMT, DPT, \#PCL     &    0.788 & 0.718 \\
    PYMT, DPT, \#PCL, ARA, SF     &    0.845 & 0.971 \\
    \bottomrule
    \end{tabularx}}
    \end{minipage}\hfill \begin{minipage}[t]{0.36\textwidth}\centering
    \resizebox{1\textwidth}{!}{
    \setlength{\tabcolsep}{5pt}
    \begin{tabularx}{1.1\textwidth}{X r r}
    \toprule
    event log & sample  & population \\
    \midrule
3. & 0.011 & 0.005\\
    6. & 0.035 & 0.071 \\
    7. & 0.152 & 0.146 \\
    8. & 1.000 & 0.952 \\
\bottomrule\\
   &\\
   &\\
\end{tabularx}}
    \end{minipage}\end{table}

However, the sample uniqueness alone does not lead to a high re-identification risk.
Therefore we also have to consider the population uniqueness
We used the method described in \Cref{sec:uniqueness_attributes} to estimate the
population uniqueness and approximate the marginals from the published event log.
In \Cref{tab:bpi18qicombi} (left), we present the average estimated
population uniqueness of five runs.
Interestingly, the population uniqueness with a single case attribute (PYMT) is already 16.1\%. Considering all case attributes, a population uniqueness of around 97\% is observed. We measure the sample uniqueness and estimate the population uniqueness for
all event logs with more than one case attribute resulting in four event logs for the analysis.
We do not consider case attributes that contain activities of the event log (i.e., the first executed activity), since we assume that an adversary does not know the exact order of executed activities.
\Cref{tab:results_qi} (right) lists the average sample uniqueness
and the average estimated population uniqueness after five runs.
We notice that not all event logs show a high uniqueness based on the case attributes. In case of the BPI Challenge 2018 event log, it can be observed that even a small number of case attributes produces a high uniqueness and thus a high re-identification risk.

\subsection{Uniqueness results based on traces}

The Sepsis event log is obtained from the information system of a Dutch hospital.
It contains events related to logistics and treatment of
patients that enter the emergency room and are suspected
to suffer from sepsis, which is a life-threatening condition that warrants immediate treatment.
Originally, the event log was analyzed regarding the adherence to guidelines on timely administration of antibiotics and, more generally, related to the overall trajectory of patients~\cite{Mannhardt.2017c}.
The data was made publicly available for research purposes~\cite{Mannhardt.2016}.
Several measures were taken to prevent identification, including:
\begin{compactitem}
    \item randomization of timestamps by perturbing the start of cases
    and adjusting timestamps of respective subsequent events accordingly
    \item pseudonymization of discharge related activities, e.g., \enquote{Release A}
    \item generalization of employee information by stating the department only
    \item pseudonymization of the working diagnosis
    \item generalization of age to groups of 5 years and at least 10 people.
\end{compactitem}
The event log consists of 1,049 cases with 16 different activities.
Each case represents the pathway through the hospital of a natural person.
The traces have an average length of $14$ points ($\min= 3$, $\max= 185$).
In contrast to the BPI Challenge 2018 event log, the Sepsis event log has only one attribute that can be used as a case attribute.

To estimate the uniqueness of traces, we use the method described in \Cref{sec:uniqueness_traces}.
The points in the Sepsis event log consist of activities, timestamps,
and departments that are currently responsible for a patient's treatment.
The \enquote{age} serves as case attribute.
Since patients are treated in different departments,
the \enquote{department} does not satisfy the time-invariant criteria of a case attribute (cf. \Cref{privacy}).
For each case, we randomly select $m$ points of the trace and count the number of traces with identical points.
In other words, we look for other traces that for example include the same activities by the same department.
We opt for a random point selection to avoid making assumptions on the adversary's knowledge.
We are aware that this may underestimate the re-identification risk.
As a consequence, a high uniqueness in our results emphasizes the re-identification risk
as a more sophisticated and optimized point selection would likely lead to an even higher uniqueness.

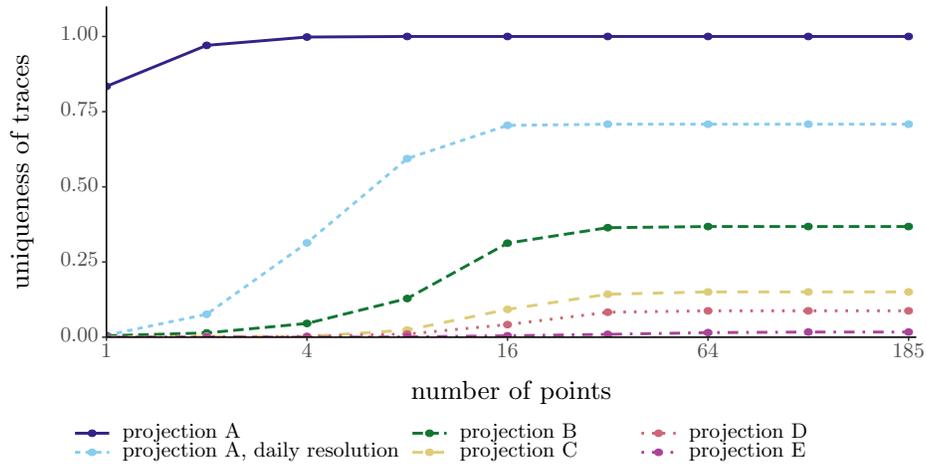
\begin{figure}[tb]
  \centering
  \begin{tikzpicture}[x=1.2pt,y=1pt,scale=0.5]
\definecolor{fillColor}{RGB}{255,255,255}
\path[use as bounding box,fill=fillColor,fill opacity=0.00] (0,0) rectangle (578.16,361.35);
\begin{scope}
\path[clip] (  0.00,  0.00) rectangle (578.16,361.35);
\definecolor{drawColor}{RGB}{255,255,255}
\definecolor{fillColor}{RGB}{255,255,255}

\path[draw=drawColor,line width= 0.6pt,line join=round,line cap=round,fill=fillColor] (  0.00,  0.00) rectangle (578.16,361.35);
\end{scope}
\begin{scope}
\path[clip] ( 62.12,105.39) rectangle (572.66,355.85);
\definecolor{fillColor}{RGB}{255,255,255}

\path[fill=fillColor] ( 62.12,105.39) rectangle (572.66,355.85);
\definecolor{drawColor}{RGB}{51,34,136}

\path[draw=drawColor,line width= 1.1pt,line join=round] ( 62.12,295.31) --
	(125.33,326.35) --
	(188.54,332.65) --
	(251.75,333.08) --
	(314.96,333.08) --
	(378.17,333.08) --
	(441.38,333.08) --
	(504.59,333.08) --
	(567.80,333.08);
\definecolor{drawColor}{RGB}{136,204,238}

\path[draw=drawColor,line width= 1.1pt,dash pattern=on 2pt off 2pt ,line join=round] ( 62.12,106.91) --
	(125.33,122.76) --
	(188.54,176.80) --
	(251.75,240.62) --
	(314.96,265.79) --
	(378.17,266.66) --
	(441.38,266.66) --
	(504.59,266.66) --
	(567.80,266.66);
\definecolor{drawColor}{RGB}{17,119,51}

\path[draw=drawColor,line width= 1.1pt,dash pattern=on 4pt off 2pt ,line join=round] ( 62.12,106.48) --
	(125.33,108.65) --
	(188.54,115.81) --
	(251.75,134.69) --
	(314.96,176.59) --
	(378.17,188.31) --
	(441.38,189.18) --
	(504.59,189.18) --
	(567.80,189.18);
\definecolor{drawColor}{RGB}{221,204,119}

\path[draw=drawColor,line width= 1.1pt,dash pattern=on 4pt off 4pt ,line join=round] ( 62.12,105.39) --
	(125.33,106.04) --
	(188.54,105.83) --
	(251.75,110.82) --
	(314.96,126.45) --
	(378.17,137.95) --
	(441.38,139.69) --
	(504.59,139.69) --
	(567.80,139.69);
\definecolor{drawColor}{RGB}{204,102,119}

\path[draw=drawColor,line width= 1.1pt,dash pattern=on 1pt off 3pt ,line join=round] ( 62.12,105.39) --
	(125.33,105.39) --
	(188.54,106.04) --
	(251.75,107.78) --
	(314.96,114.94) --
	(378.17,124.28) --
	(441.38,125.36) --
	(504.59,125.36) --
	(567.80,125.36);
\definecolor{drawColor}{RGB}{170,68,153}

\path[draw=drawColor,line width= 1.1pt,dash pattern=on 1pt off 3pt on 4pt off 3pt ,line join=round] ( 62.12,105.39) --
	(125.33,105.61) --
	(188.54,105.39) --
	(251.75,105.61) --
	(314.96,106.48) --
	(378.17,107.56) --
	(441.38,108.87) --
	(504.59,109.30) --
	(567.80,109.30);
\definecolor{drawColor}{RGB}{51,34,136}
\definecolor{fillColor}{RGB}{51,34,136}

\path[draw=drawColor,line width= 0.4pt,line join=round,line cap=round,fill=fillColor] ( 62.12,295.31) circle (  2.50);

\path[draw=drawColor,line width= 0.4pt,line join=round,line cap=round,fill=fillColor] (125.33,326.35) circle (  2.50);

\path[draw=drawColor,line width= 0.4pt,line join=round,line cap=round,fill=fillColor] (188.54,332.65) circle (  2.50);

\path[draw=drawColor,line width= 0.4pt,line join=round,line cap=round,fill=fillColor] (251.75,333.08) circle (  2.50);

\path[draw=drawColor,line width= 0.4pt,line join=round,line cap=round,fill=fillColor] (314.96,333.08) circle (  2.50);

\path[draw=drawColor,line width= 0.4pt,line join=round,line cap=round,fill=fillColor] (378.17,333.08) circle (  2.50);

\path[draw=drawColor,line width= 0.4pt,line join=round,line cap=round,fill=fillColor] (441.38,333.08) circle (  2.50);

\path[draw=drawColor,line width= 0.4pt,line join=round,line cap=round,fill=fillColor] (504.59,333.08) circle (  2.50);

\path[draw=drawColor,line width= 0.4pt,line join=round,line cap=round,fill=fillColor] (567.80,333.08) circle (  2.50);
\definecolor{drawColor}{RGB}{136,204,238}
\definecolor{fillColor}{RGB}{136,204,238}

\path[draw=drawColor,line width= 0.4pt,line join=round,line cap=round,fill=fillColor] ( 62.12,106.91) circle (  2.50);

\path[draw=drawColor,line width= 0.4pt,line join=round,line cap=round,fill=fillColor] (125.33,122.76) circle (  2.50);

\path[draw=drawColor,line width= 0.4pt,line join=round,line cap=round,fill=fillColor] (188.54,176.80) circle (  2.50);

\path[draw=drawColor,line width= 0.4pt,line join=round,line cap=round,fill=fillColor] (251.75,240.62) circle (  2.50);

\path[draw=drawColor,line width= 0.4pt,line join=round,line cap=round,fill=fillColor] (314.96,265.79) circle (  2.50);

\path[draw=drawColor,line width= 0.4pt,line join=round,line cap=round,fill=fillColor] (378.17,266.66) circle (  2.50);

\path[draw=drawColor,line width= 0.4pt,line join=round,line cap=round,fill=fillColor] (441.38,266.66) circle (  2.50);

\path[draw=drawColor,line width= 0.4pt,line join=round,line cap=round,fill=fillColor] (504.59,266.66) circle (  2.50);

\path[draw=drawColor,line width= 0.4pt,line join=round,line cap=round,fill=fillColor] (567.80,266.66) circle (  2.50);
\definecolor{drawColor}{RGB}{17,119,51}
\definecolor{fillColor}{RGB}{17,119,51}

\path[draw=drawColor,line width= 0.4pt,line join=round,line cap=round,fill=fillColor] ( 62.12,106.48) circle (  2.50);

\path[draw=drawColor,line width= 0.4pt,line join=round,line cap=round,fill=fillColor] (125.33,108.65) circle (  2.50);

\path[draw=drawColor,line width= 0.4pt,line join=round,line cap=round,fill=fillColor] (188.54,115.81) circle (  2.50);

\path[draw=drawColor,line width= 0.4pt,line join=round,line cap=round,fill=fillColor] (251.75,134.69) circle (  2.50);

\path[draw=drawColor,line width= 0.4pt,line join=round,line cap=round,fill=fillColor] (314.96,176.59) circle (  2.50);

\path[draw=drawColor,line width= 0.4pt,line join=round,line cap=round,fill=fillColor] (378.17,188.31) circle (  2.50);

\path[draw=drawColor,line width= 0.4pt,line join=round,line cap=round,fill=fillColor] (441.38,189.18) circle (  2.50);

\path[draw=drawColor,line width= 0.4pt,line join=round,line cap=round,fill=fillColor] (504.59,189.18) circle (  2.50);

\path[draw=drawColor,line width= 0.4pt,line join=round,line cap=round,fill=fillColor] (567.80,189.18) circle (  2.50);
\definecolor{drawColor}{RGB}{221,204,119}
\definecolor{fillColor}{RGB}{221,204,119}

\path[draw=drawColor,line width= 0.4pt,line join=round,line cap=round,fill=fillColor] ( 62.12,105.39) circle (  2.50);

\path[draw=drawColor,line width= 0.4pt,line join=round,line cap=round,fill=fillColor] (125.33,106.04) circle (  2.50);

\path[draw=drawColor,line width= 0.4pt,line join=round,line cap=round,fill=fillColor] (188.54,105.83) circle (  2.50);

\path[draw=drawColor,line width= 0.4pt,line join=round,line cap=round,fill=fillColor] (251.75,110.82) circle (  2.50);

\path[draw=drawColor,line width= 0.4pt,line join=round,line cap=round,fill=fillColor] (314.96,126.45) circle (  2.50);

\path[draw=drawColor,line width= 0.4pt,line join=round,line cap=round,fill=fillColor] (378.17,137.95) circle (  2.50);

\path[draw=drawColor,line width= 0.4pt,line join=round,line cap=round,fill=fillColor] (441.38,139.69) circle (  2.50);

\path[draw=drawColor,line width= 0.4pt,line join=round,line cap=round,fill=fillColor] (504.59,139.69) circle (  2.50);

\path[draw=drawColor,line width= 0.4pt,line join=round,line cap=round,fill=fillColor] (567.80,139.69) circle (  2.50);
\definecolor{drawColor}{RGB}{204,102,119}
\definecolor{fillColor}{RGB}{204,102,119}

\path[draw=drawColor,line width= 0.4pt,line join=round,line cap=round,fill=fillColor] ( 62.12,105.39) circle (  2.50);

\path[draw=drawColor,line width= 0.4pt,line join=round,line cap=round,fill=fillColor] (125.33,105.39) circle (  2.50);

\path[draw=drawColor,line width= 0.4pt,line join=round,line cap=round,fill=fillColor] (188.54,106.04) circle (  2.50);

\path[draw=drawColor,line width= 0.4pt,line join=round,line cap=round,fill=fillColor] (251.75,107.78) circle (  2.50);

\path[draw=drawColor,line width= 0.4pt,line join=round,line cap=round,fill=fillColor] (314.96,114.94) circle (  2.50);

\path[draw=drawColor,line width= 0.4pt,line join=round,line cap=round,fill=fillColor] (378.17,124.28) circle (  2.50);

\path[draw=drawColor,line width= 0.4pt,line join=round,line cap=round,fill=fillColor] (441.38,125.36) circle (  2.50);

\path[draw=drawColor,line width= 0.4pt,line join=round,line cap=round,fill=fillColor] (504.59,125.36) circle (  2.50);

\path[draw=drawColor,line width= 0.4pt,line join=round,line cap=round,fill=fillColor] (567.80,125.36) circle (  2.50);
\definecolor{drawColor}{RGB}{170,68,153}
\definecolor{fillColor}{RGB}{170,68,153}

\path[draw=drawColor,line width= 0.4pt,line join=round,line cap=round,fill=fillColor] ( 62.12,105.39) circle (  2.50);

\path[draw=drawColor,line width= 0.4pt,line join=round,line cap=round,fill=fillColor] (125.33,105.61) circle (  2.50);

\path[draw=drawColor,line width= 0.4pt,line join=round,line cap=round,fill=fillColor] (188.54,105.39) circle (  2.50);

\path[draw=drawColor,line width= 0.4pt,line join=round,line cap=round,fill=fillColor] (251.75,105.61) circle (  2.50);

\path[draw=drawColor,line width= 0.4pt,line join=round,line cap=round,fill=fillColor] (314.96,106.48) circle (  2.50);

\path[draw=drawColor,line width= 0.4pt,line join=round,line cap=round,fill=fillColor] (378.17,107.56) circle (  2.50);

\path[draw=drawColor,line width= 0.4pt,line join=round,line cap=round,fill=fillColor] (441.38,108.87) circle (  2.50);

\path[draw=drawColor,line width= 0.4pt,line join=round,line cap=round,fill=fillColor] (504.59,109.30) circle (  2.50);

\path[draw=drawColor,line width= 0.4pt,line join=round,line cap=round,fill=fillColor] (567.80,109.30) circle (  2.50);
\end{scope}
\begin{scope}
\path[clip] (  0.00,  0.00) rectangle (578.16,361.35);
\definecolor{drawColor}{RGB}{0,0,0}

\path[draw=drawColor,line width= 0.6pt,line join=round] ( 62.12,105.39) --
	( 62.12,355.85);
\end{scope}
\begin{scope}
\path[clip] (  0.00,  0.00) rectangle (578.16,361.35);
\definecolor{drawColor}{gray}{0.30}

\node[text=drawColor,anchor=base east,inner sep=0pt, outer sep=0pt, scale=  0.88] at ( 57.17,102.36) {0.00};

\node[text=drawColor,anchor=base east,inner sep=0pt, outer sep=0pt, scale=  0.88] at ( 57.17,159.28) {0.25};

\node[text=drawColor,anchor=base east,inner sep=0pt, outer sep=0pt, scale=  0.88] at ( 57.17,216.21) {0.50};

\node[text=drawColor,anchor=base east,inner sep=0pt, outer sep=0pt, scale=  0.88] at ( 57.17,273.13) {0.75};

\node[text=drawColor,anchor=base east,inner sep=0pt, outer sep=0pt, scale=  0.88] at ( 57.17,330.05) {1.00};
\end{scope}
\begin{scope}
\path[clip] (  0.00,  0.00) rectangle (578.16,361.35);
\definecolor{drawColor}{gray}{0.20}

\path[draw=drawColor,line width= 0.6pt,line join=round] ( 59.37,105.39) --
	( 62.12,105.39);

\path[draw=drawColor,line width= 0.6pt,line join=round] ( 59.37,162.31) --
	( 62.12,162.31);

\path[draw=drawColor,line width= 0.6pt,line join=round] ( 59.37,219.24) --
	( 62.12,219.24);

\path[draw=drawColor,line width= 0.6pt,line join=round] ( 59.37,276.16) --
	( 62.12,276.16);

\path[draw=drawColor,line width= 0.6pt,line join=round] ( 59.37,333.08) --
	( 62.12,333.08);
\end{scope}
\begin{scope}
\path[clip] (  0.00,  0.00) rectangle (578.16,361.35);
\definecolor{drawColor}{RGB}{0,0,0}

\path[draw=drawColor,line width= 0.6pt,line join=round] ( 62.12,105.39) --
	(572.66,105.39);
\end{scope}
\begin{scope}
\path[clip] (  0.00,  0.00) rectangle (578.16,361.35);
\definecolor{drawColor}{gray}{0.20}

\path[draw=drawColor,line width= 0.6pt,line join=round] ( 62.12,102.64) --
	( 62.12,105.39);

\path[draw=drawColor,line width= 0.6pt,line join=round] (188.54,102.64) --
	(188.54,105.39);

\path[draw=drawColor,line width= 0.6pt,line join=round] (314.96,102.64) --
	(314.96,105.39);

\path[draw=drawColor,line width= 0.6pt,line join=round] (441.38,102.64) --
	(441.38,105.39);

\path[draw=drawColor,line width= 0.6pt,line join=round] (567.80,102.64) --
	(567.80,105.39);
\end{scope}
\begin{scope}
\path[clip] (  0.00,  0.00) rectangle (578.16,361.35);
\definecolor{drawColor}{gray}{0.30}

\node[text=drawColor,anchor=base,inner sep=0pt, outer sep=0pt, scale=  0.88] at ( 62.12, 90) {1};

\node[text=drawColor,anchor=base,inner sep=0pt, outer sep=0pt, scale=  0.88] at (188.54, 90) {4};

\node[text=drawColor,anchor=base,inner sep=0pt, outer sep=0pt, scale=  0.88] at (314.96, 90) {16};

\node[text=drawColor,anchor=base,inner sep=0pt, outer sep=0pt, scale=  0.88] at (441.38, 90) {64};

\node[text=drawColor,anchor=base,inner sep=0pt, outer sep=0pt, scale=  0.88] at (567.80, 90) {185};
\end{scope}
\begin{scope}
\path[clip] (  0.00,  0.00) rectangle (578.16,361.35);
\definecolor{drawColor}{RGB}{0,0,0}

\node[text=drawColor,anchor=base,inner sep=0pt, outer sep=0pt, scale=  1.10] at (317.39, 82.11) {};

\node[text=drawColor,anchor=base,inner sep=0pt, outer sep=0pt, scale=  1.10] at (317.39, 70.23) {};

\node[text=drawColor,anchor=base,inner sep=0pt, outer sep=0pt, scale=  1.10] at (317.39, 58.35) {number of points};
\end{scope}
\begin{scope}
\path[clip] (  0.00,  0.00) rectangle (578.16,361.35);
\definecolor{drawColor}{RGB}{0,0,0}

\node[text=drawColor,rotate= 90.00,anchor=base,inner sep=0pt, outer sep=0pt, scale=  1.10] at ( 13.08,230.62) {uniqueness of traces};

\node[text=drawColor,rotate= 90.00,anchor=base,inner sep=0pt, outer sep=0pt, scale=  1.10] at ( 24.96,230.62) {};

\node[text=drawColor,rotate= 90.00,anchor=base,inner sep=0pt, outer sep=0pt, scale=  1.10] at ( 36.84,230.62) {};
\end{scope}
\begin{scope}
\path[clip] (  0.00,  0.00) rectangle (578.16,361.35);
\definecolor{drawColor}{RGB}{51,34,136}

\path[draw=drawColor,line width= 1.1pt,line join=round] (42.30, 32.68) -- (63.86, 32.68);
\end{scope}
\begin{scope}
\path[clip] (  0.00,  0.00) rectangle (578.16,361.35);
\definecolor{drawColor}{RGB}{51,34,136}
\definecolor{fillColor}{RGB}{51,34,136}

\path[draw=drawColor,line width= 0.4pt,line join=round,line cap=round,fill=fillColor] (53.08, 32.68) circle (  2.50);
\end{scope}
\begin{scope}
\path[clip] (  0.00,  0.00) rectangle (578.16,361.35);
\definecolor{drawColor}{RGB}{136,204,238}

\path[draw=drawColor,line width= 1.1pt,dash pattern=on 2pt off 2pt ,line join=round] (42.30, 18.23) -- (63.86, 18.23);
\end{scope}
\begin{scope}
\path[clip] (  0.00,  0.00) rectangle (578.16,361.35);
\definecolor{drawColor}{RGB}{136,204,238}
\definecolor{fillColor}{RGB}{136,204,238}

\path[draw=drawColor,line width= 0.4pt,line join=round,line cap=round,fill=fillColor] (53.08, 18.23) circle (  2.50);
\end{scope}
\begin{scope}
\path[clip] (  0.00,  0.00) rectangle (578.16,361.35);
\definecolor{drawColor}{RGB}{17,119,51}

\path[draw=drawColor,line width= 1.1pt,dash pattern=on 4pt off 2pt ,line join=round] (255.01, 32.68) -- (276.58, 32.68);
\end{scope}
\begin{scope}
\path[clip] (  0.00,  0.00) rectangle (578.16,361.35);
\definecolor{drawColor}{RGB}{17,119,51}
\definecolor{fillColor}{RGB}{17,119,51}

\path[draw=drawColor,line width= 0.4pt,line join=round,line cap=round,fill=fillColor] (265.79, 32.68) circle (  2.50);
\end{scope}
\begin{scope}
\path[clip] (  0.00,  0.00) rectangle (578.16,361.35);
\definecolor{drawColor}{RGB}{221,204,119}

\path[draw=drawColor,line width= 1.1pt,dash pattern=on 4pt off 4pt ,line join=round] (255.01, 18.23) -- (276.58, 18.23);
\end{scope}
\begin{scope}
\path[clip] (  0.00,  0.00) rectangle (578.16,361.35);
\definecolor{drawColor}{RGB}{221,204,119}
\definecolor{fillColor}{RGB}{221,204,119}

\path[draw=drawColor,line width= 0.4pt,line join=round,line cap=round,fill=fillColor] (265.79, 18.23) circle (  2.50);
\end{scope}
\begin{scope}
\path[clip] (  0.00,  0.00) rectangle (578.16,361.35);
\definecolor{drawColor}{RGB}{204,102,119}

\path[draw=drawColor,line width= 1.1pt,dash pattern=on 1pt off 3pt ,line join=round] (399.41, 32.68) -- (420.97, 32.68);
\end{scope}
\begin{scope}
\path[clip] (  0.00,  0.00) rectangle (578.16,361.35);
\definecolor{drawColor}{RGB}{204,102,119}
\definecolor{fillColor}{RGB}{204,102,119}

\path[draw=drawColor,line width= 0.4pt,line join=round,line cap=round,fill=fillColor] (410.19, 32.68) circle (  2.50);
\end{scope}
\begin{scope}
\path[clip] (  0.00,  0.00) rectangle (578.16,361.35);
\definecolor{drawColor}{RGB}{170,68,153}

\path[draw=drawColor,line width= 1.1pt,dash pattern=on 1pt off 3pt on 4pt off 3pt ,line join=round] (399.41, 18.23) -- (420.97, 18.23);
\end{scope}
\begin{scope}
\path[clip] (  0.00,  0.00) rectangle (478.16,361.35);
\definecolor{drawColor}{RGB}{170,68,153}
\definecolor{fillColor}{RGB}{170,68,153}

\path[draw=drawColor,line width= 0.4pt,line join=round,line cap=round,fill=fillColor] (410.19, 18.23) circle (  2.50);
\end{scope}
\begin{scope}
\path[clip] (  0.00,  0.00) rectangle (578.16,361.35);
\definecolor{drawColor}{RGB}{0,0,0}

\node[text=drawColor,anchor=base west,inner sep=0pt, outer sep=0pt, scale=  0.88] at (72.21, 29.65) {projection A};
\end{scope}
\begin{scope}
\path[clip] (  0.00,  0.00) rectangle (578.16,361.35);
\definecolor{drawColor}{RGB}{0,0,0}

\node[text=drawColor,anchor=base west,inner sep=0pt, outer sep=0pt, scale=  0.88] at (72.21, 15.20) {projection A, daily resolution};
\end{scope}
\begin{scope}
\path[clip] (  0.00,  0.00) rectangle (578.16,361.35);
\definecolor{drawColor}{RGB}{0,0,0}

\node[text=drawColor,anchor=base west,inner sep=0pt, outer sep=0pt, scale=  0.88] at (284.93, 29.65) {projection B};
\end{scope}
\begin{scope}
\path[clip] (  0.00,  0.00) rectangle (578.16,361.35);
\definecolor{drawColor}{RGB}{0,0,0}

\node[text=drawColor,anchor=base west,inner sep=0pt, outer sep=0pt, scale=  0.88] at (284.93, 15.20) {projection C};
\end{scope}
\begin{scope}
\path[clip] (  0.00,  0.00) rectangle (578.16,361.35);
\definecolor{drawColor}{RGB}{0,0,0}

\node[text=drawColor,anchor=base west,inner sep=0pt, outer sep=0pt, scale=  0.88] at (429.32, 29.65) {projection D};
\end{scope}
\begin{scope}
\path[clip] (  0.00,  0.00) rectangle (578.16,361.35);
\definecolor{drawColor}{RGB}{0,0,0}

\node[text=drawColor,anchor=base west,inner sep=0pt, outer sep=0pt, scale=  0.88] at (429.32, 15.20) {projection E};
\end{scope}
\end{tikzpicture}
   \vspace*{-1em}
  \caption{Uniqueness based on traces for Sepsis event log.}
  \label{fig:unicity_sepsis}
\end{figure}

In \Cref{fig:unicity_sepsis}, we show the uniqueness of traces for different values of $m$ points
and different projections.\footnote{Code available at 
\url{https://github.com/d-o-m-i-n-i-k/re-identification-risk}.} As expected, we 
generally observe that more points lead to a higher uniqueness. Assuming that 
timestamps are correct (which they are not),
projection~A shows that four points including the activity and the timestamp are sufficient to identify all traces. By generalizing timestamps, i.e., reducing the resolution to days, only 31\% of traces are unique when considering four points and 70\% when considering all points of a trace.
Hence, the results clearly show the impact of generalization on the re-identification risk.

The privacy-enhancing effect of removing values from the event log becomes apparent,
when considering the other projections. Projection~B, for example, omits timestamps
but otherwise assumes that an adversary has background knowledge on all activities, case and event attributes. Yet, it is able to significantly limit the uniqueness to approximately 37\%.
Projection~D, where case attributes and activities are still included,
is even able to limit the uniqueness of traces to a maximum of 9\%.
The uniqueness of traces remains stable for more than 64 points since only 2\% 
of the traces have more than 64 points. 

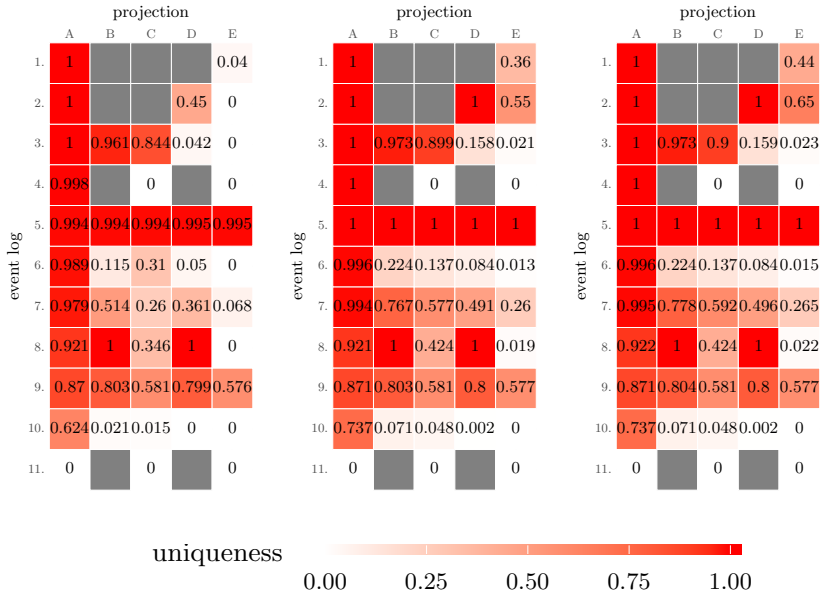
\begin{figure}[tb]
\begin{minipage}[t]{\textwidth}
  \begin{minipage}[t]{0.3\textwidth}
  \resizebox{1\textwidth}{!}{
    \begin{tikzpicture}[x=1pt,y=1pt]
\definecolor{fillColor}{RGB}{255,255,255}
\path[use as bounding box,fill=fillColor,fill opacity=0.00] (0,0) rectangle (289.08,578.16);
\begin{scope}
\path[clip] ( 55.60, 55.46) rectangle (273.82,535.56);
\definecolor{drawColor}{gray}{0.92}

\path[draw=drawColor,line width= 0.6pt,line join=round] ( 55.60, 77.28) --
	(273.82, 77.28);

\path[draw=drawColor,line width= 0.6pt,line join=round] ( 55.60,120.93) --
	(273.82,120.93);

\path[draw=drawColor,line width= 0.6pt,line join=round] ( 55.60,164.57) --
	(273.82,164.57);

\path[draw=drawColor,line width= 0.6pt,line join=round] ( 55.60,208.22) --
	(273.82,208.22);

\path[draw=drawColor,line width= 0.6pt,line join=round] ( 55.60,251.86) --
	(273.82,251.86);

\path[draw=drawColor,line width= 0.6pt,line join=round] ( 55.60,295.51) --
	(273.82,295.51);

\path[draw=drawColor,line width= 0.6pt,line join=round] ( 55.60,339.15) --
	(273.82,339.15);

\path[draw=drawColor,line width= 0.6pt,line join=round] ( 55.60,382.80) --
	(273.82,382.80);

\path[draw=drawColor,line width= 0.6pt,line join=round] ( 55.60,426.44) --
	(273.82,426.44);

\path[draw=drawColor,line width= 0.6pt,line join=round] ( 55.60,470.09) --
	(273.82,470.09);

\path[draw=drawColor,line width= 0.6pt,line join=round] ( 55.60,513.73) --
	(273.82,513.73);

\path[draw=drawColor,line width= 0.6pt,line join=round] ( 77.42, 55.46) --
	( 77.42,535.56);

\path[draw=drawColor,line width= 0.6pt,line join=round] (121.07, 55.46) --
	(121.07,535.56);

\path[draw=drawColor,line width= 0.6pt,line join=round] (164.71, 55.46) --
	(164.71,535.56);

\path[draw=drawColor,line width= 0.6pt,line join=round] (208.36, 55.46) --
	(208.36,535.56);

\path[draw=drawColor,line width= 0.6pt,line join=round] (252.00, 55.46) --
	(252.00,535.56);
\definecolor{drawColor}{RGB}{255,255,255}
\definecolor{fillColor}{RGB}{255,0,0}

\path[draw=drawColor,line width= 0.1pt,line join=round,fill=fillColor] ( 55.60,491.91) rectangle ( 99.24,535.56);
\definecolor{fillColor}{RGB}{255,54,28}

\path[draw=drawColor,line width= 0.1pt,line join=round,fill=fillColor] ( 55.60,186.39) rectangle ( 99.24,230.04);
\definecolor{fillColor}{RGB}{255,24,8}

\path[draw=drawColor,line width= 0.1pt,line join=round,fill=fillColor] ( 55.60,230.04) rectangle ( 99.24,273.68);
\definecolor{fillColor}{RGB}{255,0,0}

\path[draw=drawColor,line width= 0.1pt,line join=round,fill=fillColor] ( 55.60,404.62) rectangle ( 99.24,448.27);

\path[draw=drawColor,line width= 0.1pt,line join=round,fill=fillColor] ( 55.60,448.27) rectangle ( 99.24,491.91);
\definecolor{fillColor}{RGB}{255,3,1}

\path[draw=drawColor,line width= 0.1pt,line join=round,fill=fillColor] ( 55.60,360.97) rectangle ( 99.24,404.62);
\definecolor{fillColor}{RGB}{255,255,255}

\path[draw=drawColor,line width= 0.1pt,line join=round,fill=fillColor] ( 55.60, 55.46) rectangle ( 99.24, 99.10);
\definecolor{fillColor}{RGB}{255,72,41}

\path[draw=drawColor,line width= 0.1pt,line join=round,fill=fillColor] ( 55.60,142.75) rectangle ( 99.24,186.39);
\definecolor{fillColor}{RGB}{255,15,4}

\path[draw=drawColor,line width= 0.1pt,line join=round,fill=fillColor] ( 55.60,273.68) rectangle ( 99.24,317.33);
\definecolor{fillColor}{RGB}{255,132,100}

\path[draw=drawColor,line width= 0.1pt,line join=round,fill=fillColor] ( 55.60, 99.10) rectangle ( 99.24,142.75);
\definecolor{fillColor}{RGB}{255,8,2}

\path[draw=drawColor,line width= 0.1pt,line join=round,fill=fillColor] ( 55.60,317.33) rectangle ( 99.24,360.97);
\definecolor{fillColor}{gray}{0.50}

\path[draw=drawColor,line width= 0.1pt,line join=round,fill=fillColor] ( 99.24,491.91) rectangle (142.89,535.56);
\definecolor{fillColor}{RGB}{255,0,0}

\path[draw=drawColor,line width= 0.1pt,line join=round,fill=fillColor] ( 99.24,186.39) rectangle (142.89,230.04);
\definecolor{fillColor}{RGB}{255,155,126}

\path[draw=drawColor,line width= 0.1pt,line join=round,fill=fillColor] ( 99.24,230.04) rectangle (142.89,273.68);
\definecolor{fillColor}{RGB}{255,36,15}

\path[draw=drawColor,line width= 0.1pt,line join=round,fill=fillColor] ( 99.24,404.62) rectangle (142.89,448.27);
\definecolor{fillColor}{gray}{0.50}

\path[draw=drawColor,line width= 0.1pt,line join=round,fill=fillColor] ( 99.24,448.27) rectangle (142.89,491.91);

\path[draw=drawColor,line width= 0.1pt,line join=round,fill=fillColor] ( 99.24,360.97) rectangle (142.89,404.62);

\path[draw=drawColor,line width= 0.1pt,line join=round,fill=fillColor] ( 99.24, 55.46) rectangle (142.89, 99.10);
\definecolor{fillColor}{RGB}{255,90,58}

\path[draw=drawColor,line width= 0.1pt,line join=round,fill=fillColor] ( 99.24,142.75) rectangle (142.89,186.39);
\definecolor{fillColor}{RGB}{255,233,225}

\path[draw=drawColor,line width= 0.1pt,line join=round,fill=fillColor] ( 99.24,273.68) rectangle (142.89,317.33);
\definecolor{fillColor}{RGB}{255,251,249}

\path[draw=drawColor,line width= 0.1pt,line join=round,fill=fillColor] ( 99.24, 99.10) rectangle (142.89,142.75);
\definecolor{fillColor}{RGB}{255,8,2}

\path[draw=drawColor,line width= 0.1pt,line join=round,fill=fillColor] ( 99.24,317.33) rectangle (142.89,360.97);
\definecolor{fillColor}{gray}{0.50}

\path[draw=drawColor,line width= 0.1pt,line join=round,fill=fillColor] (142.89,491.91) rectangle (186.53,535.56);
\definecolor{fillColor}{RGB}{255,189,167}

\path[draw=drawColor,line width= 0.1pt,line join=round,fill=fillColor] (142.89,186.39) rectangle (186.53,230.04);
\definecolor{fillColor}{RGB}{255,205,188}

\path[draw=drawColor,line width= 0.1pt,line join=round,fill=fillColor] (142.89,230.04) rectangle (186.53,273.68);
\definecolor{fillColor}{RGB}{255,79,48}

\path[draw=drawColor,line width= 0.1pt,line join=round,fill=fillColor] (142.89,404.62) rectangle (186.53,448.27);
\definecolor{fillColor}{gray}{0.50}

\path[draw=drawColor,line width= 0.1pt,line join=round,fill=fillColor] (142.89,448.27) rectangle (186.53,491.91);
\definecolor{fillColor}{RGB}{255,255,255}

\path[draw=drawColor,line width= 0.1pt,line join=round,fill=fillColor] (142.89,360.97) rectangle (186.53,404.62);

\path[draw=drawColor,line width= 0.1pt,line join=round,fill=fillColor] (142.89, 55.46) rectangle (186.53, 99.10);
\definecolor{fillColor}{RGB}{255,141,110}

\path[draw=drawColor,line width= 0.1pt,line join=round,fill=fillColor] (142.89,142.75) rectangle (186.53,186.39);
\definecolor{fillColor}{RGB}{255,196,176}

\path[draw=drawColor,line width= 0.1pt,line join=round,fill=fillColor] (142.89,273.68) rectangle (186.53,317.33);
\definecolor{fillColor}{RGB}{255,252,251}

\path[draw=drawColor,line width= 0.1pt,line join=round,fill=fillColor] (142.89, 99.10) rectangle (186.53,142.75);
\definecolor{fillColor}{RGB}{255,9,2}

\path[draw=drawColor,line width= 0.1pt,line join=round,fill=fillColor] (142.89,317.33) rectangle (186.53,360.97);
\definecolor{fillColor}{gray}{0.50}

\path[draw=drawColor,line width= 0.1pt,line join=round,fill=fillColor] (186.53,491.91) rectangle (230.18,535.56);
\definecolor{fillColor}{RGB}{255,0,0}

\path[draw=drawColor,line width= 0.1pt,line join=round,fill=fillColor] (186.53,186.39) rectangle (230.18,230.04);
\definecolor{fillColor}{RGB}{255,186,163}

\path[draw=drawColor,line width= 0.1pt,line join=round,fill=fillColor] (186.53,230.04) rectangle (230.18,273.68);
\definecolor{fillColor}{RGB}{255,247,244}

\path[draw=drawColor,line width= 0.1pt,line join=round,fill=fillColor] (186.53,404.62) rectangle (230.18,448.27);
\definecolor{fillColor}{RGB}{255,168,141}

\path[draw=drawColor,line width= 0.1pt,line join=round,fill=fillColor] (186.53,448.27) rectangle (230.18,491.91);
\definecolor{fillColor}{gray}{0.50}

\path[draw=drawColor,line width= 0.1pt,line join=round,fill=fillColor] (186.53,360.97) rectangle (230.18,404.62);

\path[draw=drawColor,line width= 0.1pt,line join=round,fill=fillColor] (186.53, 55.46) rectangle (230.18, 99.10);
\definecolor{fillColor}{RGB}{255,92,59}

\path[draw=drawColor,line width= 0.1pt,line join=round,fill=fillColor] (186.53,142.75) rectangle (230.18,186.39);
\definecolor{fillColor}{RGB}{255,246,242}

\path[draw=drawColor,line width= 0.1pt,line join=round,fill=fillColor] (186.53,273.68) rectangle (230.18,317.33);
\definecolor{fillColor}{RGB}{255,255,255}

\path[draw=drawColor,line width= 0.1pt,line join=round,fill=fillColor] (186.53, 99.10) rectangle (230.18,142.75);
\definecolor{fillColor}{RGB}{255,7,2}

\path[draw=drawColor,line width= 0.1pt,line join=round,fill=fillColor] (186.53,317.33) rectangle (230.18,360.97);
\definecolor{fillColor}{RGB}{255,247,245}

\path[draw=drawColor,line width= 0.1pt,line join=round,fill=fillColor] (230.18,491.91) rectangle (273.82,535.56);
\definecolor{fillColor}{RGB}{255,255,255}

\path[draw=drawColor,line width= 0.1pt,line join=round,fill=fillColor] (230.18,186.39) rectangle (273.82,230.04);
\definecolor{fillColor}{RGB}{255,242,237}

\path[draw=drawColor,line width= 0.1pt,line join=round,fill=fillColor] (230.18,230.04) rectangle (273.82,273.68);
\definecolor{fillColor}{RGB}{255,255,255}

\path[draw=drawColor,line width= 0.1pt,line join=round,fill=fillColor] (230.18,404.62) rectangle (273.82,448.27);

\path[draw=drawColor,line width= 0.1pt,line join=round,fill=fillColor] (230.18,448.27) rectangle (273.82,491.91);

\path[draw=drawColor,line width= 0.1pt,line join=round,fill=fillColor] (230.18,360.97) rectangle (273.82,404.62);

\path[draw=drawColor,line width= 0.1pt,line join=round,fill=fillColor] (230.18, 55.46) rectangle (273.82, 99.10);
\definecolor{fillColor}{RGB}{255,142,111}

\path[draw=drawColor,line width= 0.1pt,line join=round,fill=fillColor] (230.18,142.75) rectangle (273.82,186.39);
\definecolor{fillColor}{RGB}{255,255,255}

\path[draw=drawColor,line width= 0.1pt,line join=round,fill=fillColor] (230.18,273.68) rectangle (273.82,317.33);

\path[draw=drawColor,line width= 0.1pt,line join=round,fill=fillColor] (230.18, 99.10) rectangle (273.82,142.75);
\definecolor{fillColor}{RGB}{255,7,2}

\path[draw=drawColor,line width= 0.1pt,line join=round,fill=fillColor] (230.18,317.33) rectangle (273.82,360.97);
\definecolor{drawColor}{RGB}{0,0,0}

\node[text=drawColor,anchor=base,inner sep=0pt, outer sep=0pt, scale=  2] at ( 77.42,509.93) {1};

\node[text=drawColor,anchor=base,inner sep=0pt, outer sep=0pt, scale=  2] at ( 77.42,204.41) {0.921};

\node[text=drawColor,anchor=base,inner sep=0pt, outer sep=0pt, scale=  2] at ( 77.42,248.06) {0.979};

\node[text=drawColor,anchor=base,inner sep=0pt, outer sep=0pt, scale=  2] at ( 77.42,422.64) {1};

\node[text=drawColor,anchor=base,inner sep=0pt, outer sep=0pt, scale=  2] at ( 77.42,466.29) {1};

\node[text=drawColor,anchor=base,inner sep=0pt, outer sep=0pt, scale=  2] at ( 77.42,379.00) {0.998};

\node[text=drawColor,anchor=base,inner sep=0pt, outer sep=0pt, scale=  2] at ( 77.42, 73.48) {0};

\node[text=drawColor,anchor=base,inner sep=0pt, outer sep=0pt, scale=  2] at ( 77.42,160.77) {0.87};

\node[text=drawColor,anchor=base,inner sep=0pt, outer sep=0pt, scale=  2] at ( 77.42,291.71) {0.989};

\node[text=drawColor,anchor=base,inner sep=0pt, outer sep=0pt, scale=  2] at ( 77.42,117.12) {0.624};

\node[text=drawColor,anchor=base,inner sep=0pt, outer sep=0pt, scale=  2] at ( 77.42,335.35) {0.994};

\node[text=drawColor,anchor=base,inner sep=0pt, outer sep=0pt, scale=  2] at (121.07,204.41) {1};

\node[text=drawColor,anchor=base,inner sep=0pt, outer sep=0pt, scale=  2] at (121.07,248.06) {0.514};

\node[text=drawColor,anchor=base,inner sep=0pt, outer sep=0pt, scale=  2] at (121.07,422.64) {0.961};

\node[text=drawColor,anchor=base,inner sep=0pt, outer sep=0pt, scale=  2] at (121.07,160.77) {0.803};

\node[text=drawColor,anchor=base,inner sep=0pt, outer sep=0pt, scale=  2] at (121.07,291.71) {0.115};

\node[text=drawColor,anchor=base,inner sep=0pt, outer sep=0pt, scale=  2] at (121.07,117.12) {0.021};

\node[text=drawColor,anchor=base,inner sep=0pt, outer sep=0pt, scale=  2] at (121.07,335.35) {0.994};

\node[text=drawColor,anchor=base,inner sep=0pt, outer sep=0pt, scale=  2] at (164.71,204.41) {0.346};

\node[text=drawColor,anchor=base,inner sep=0pt, outer sep=0pt, scale=  2] at (164.71,248.06) {0.26};

\node[text=drawColor,anchor=base,inner sep=0pt, outer sep=0pt, scale=  2] at (164.71,422.64) {0.844};

\node[text=drawColor,anchor=base,inner sep=0pt, outer sep=0pt, scale=  2] at (164.71,379.00) {0};

\node[text=drawColor,anchor=base,inner sep=0pt, outer sep=0pt, scale=  2] at (164.71, 73.48) {0};

\node[text=drawColor,anchor=base,inner sep=0pt, outer sep=0pt, scale=  2] at (164.71,160.77) {0.581};

\node[text=drawColor,anchor=base,inner sep=0pt, outer sep=0pt, scale=  2] at (164.71,291.71) {0.31};

\node[text=drawColor,anchor=base,inner sep=0pt, outer sep=0pt, scale=  2] at (164.71,117.12) {0.015};

\node[text=drawColor,anchor=base,inner sep=0pt, outer sep=0pt, scale=  2] at (164.71,335.35) {0.994};

\node[text=drawColor,anchor=base,inner sep=0pt, outer sep=0pt, scale=  2] at (208.36,204.41) {1};

\node[text=drawColor,anchor=base,inner sep=0pt, outer sep=0pt, scale=  2] at (208.36,248.06) {0.361};

\node[text=drawColor,anchor=base,inner sep=0pt, outer sep=0pt, scale=  2] at (208.36,422.64) {0.042};

\node[text=drawColor,anchor=base,inner sep=0pt, outer sep=0pt, scale=  2] at (208.36,466.29) {0.45};

\node[text=drawColor,anchor=base,inner sep=0pt, outer sep=0pt, scale=  2] at (208.36,160.77) {0.799};

\node[text=drawColor,anchor=base,inner sep=0pt, outer sep=0pt, scale=  2] at (208.36,291.71) {0.05};

\node[text=drawColor,anchor=base,inner sep=0pt, outer sep=0pt, scale=  2] at (208.36,117.12) {0};

\node[text=drawColor,anchor=base,inner sep=0pt, outer sep=0pt, scale=  2] at (208.36,335.35) {0.995};

\node[text=drawColor,anchor=base,inner sep=0pt, outer sep=0pt, scale=  2] at (252.00,509.93) {0.04};

\node[text=drawColor,anchor=base,inner sep=0pt, outer sep=0pt, scale=  2] at (252.00,204.41) {0};

\node[text=drawColor,anchor=base,inner sep=0pt, outer sep=0pt, scale=  2] at (252.00,248.06) {0.068};

\node[text=drawColor,anchor=base,inner sep=0pt, outer sep=0pt, scale=  2] at (252.00,422.64) {0};

\node[text=drawColor,anchor=base,inner sep=0pt, outer sep=0pt, scale=  2] at (252.00,466.29) {0};

\node[text=drawColor,anchor=base,inner sep=0pt, outer sep=0pt, scale=  2] at (252.00,379.00) {0};

\node[text=drawColor,anchor=base,inner sep=0pt, outer sep=0pt, scale=  2] at (252.00, 73.48) {0};

\node[text=drawColor,anchor=base,inner sep=0pt, outer sep=0pt, scale=  2] at (252.00,160.77) {0.576};

\node[text=drawColor,anchor=base,inner sep=0pt, outer sep=0pt, scale=  2] at (252.00,291.71) {0};

\node[text=drawColor,anchor=base,inner sep=0pt, outer sep=0pt, scale=  2] at (252.00,117.12) {0};

\node[text=drawColor,anchor=base,inner sep=0pt, outer sep=0pt, scale=  2] at (252.00,335.35) {0.995};
\end{scope}
\begin{scope}
\path[clip] (  0.00,  0.00) rectangle (289.08,578.16);
\definecolor{drawColor}{gray}{0.30}

\node[text=drawColor,anchor=base,inner sep=0pt, outer sep=0pt, scale=  1.5] at ( 77.42,540.51) {A};

\node[text=drawColor,anchor=base,inner sep=0pt, outer sep=0pt, scale=  1.5] at (121.07,540.51) {B};

\node[text=drawColor,anchor=base,inner sep=0pt, outer sep=0pt, scale=  1.5] at (164.71,540.51) {C};

\node[text=drawColor,anchor=base,inner sep=0pt, outer sep=0pt, scale=  1.5] at (208.36,540.51) {D};

\node[text=drawColor,anchor=base,inner sep=0pt, outer sep=0pt, scale=  1.5] at (252.00,540.51) {E};
\end{scope}
\begin{scope}
\path[clip] (  0.00,  0.00) rectangle (289.08,578.16);
\definecolor{drawColor}{gray}{0.30}

\node[text=drawColor,anchor=base east,inner sep=0pt, outer sep=0pt, scale=  1.5] at ( 50.65, 74.25) {11.};

\node[text=drawColor,anchor=base east,inner sep=0pt, outer sep=0pt, scale=  1.5] at ( 50.65,117.90) {10.};

\node[text=drawColor,anchor=base east,inner sep=0pt, outer sep=0pt, scale=  1.5] at ( 50.65,161.54) {9.};

\node[text=drawColor,anchor=base east,inner sep=0pt, outer sep=0pt, scale=  1.5] at ( 50.65,205.19) {8.};

\node[text=drawColor,anchor=base east,inner sep=0pt, outer sep=0pt, scale=  1.5] at ( 50.65,248.83) {7.};

\node[text=drawColor,anchor=base east,inner sep=0pt, outer sep=0pt, scale=  1.5] at ( 50.65,292.48) {6.};

\node[text=drawColor,anchor=base east,inner sep=0pt, outer sep=0pt, scale=  1.5] at ( 50.65,336.12) {5.};

\node[text=drawColor,anchor=base east,inner sep=0pt, outer sep=0pt, scale=  1.5] at ( 50.65,379.77) {4.};

\node[text=drawColor,anchor=base east,inner sep=0pt, outer sep=0pt, scale=  1.5] at ( 50.65,423.41) {3.};

\node[text=drawColor,anchor=base east,inner sep=0pt, outer sep=0pt, scale=  1.5] at ( 50.65,467.06) {2.};

\node[text=drawColor,anchor=base east,inner sep=0pt, outer sep=0pt, scale=  1.5] at ( 50.65,510.70) {1.};
\end{scope}
\begin{scope}
\path[clip] (  0.00,  0.00) rectangle (289.08,578.16);
\definecolor{drawColor}{RGB}{0,0,0}

\node[text=drawColor,anchor=base,inner sep=0pt, outer sep=0pt, scale=  2] at (164.71,563.14) {projection};

\node[text=drawColor,anchor=base,inner sep=0pt, outer sep=0pt, scale=  2] at (164.71,551.26) {};
\end{scope}
\begin{scope}
\path[clip] (  0.00,  0.00) rectangle (289.08,578.16);
\definecolor{drawColor}{RGB}{0,0,0}

\node[text=drawColor,rotate= 90.00,anchor=base,inner sep=0pt, outer sep=0pt, scale=  2] at ( 22.83,295.51) {event log};

\node[text=drawColor,rotate= 90.00,anchor=base,inner sep=0pt, outer sep=0pt, scale=  2] at ( 34.71,295.51) {};
\end{scope}
\begin{scope}
\path[clip] (  0.00,  0.00) rectangle (289.08,578.16);
\definecolor{drawColor}{RGB}{255,255,255}

\path[draw=drawColor,line width= 0.2pt,line join=round] (137.24, 24.50) -- (137.24, 27.40);

\path[draw=drawColor,line width= 0.2pt,line join=round] (171.02, 24.50) -- (171.02, 27.40);

\path[draw=drawColor,line width= 0.2pt,line join=round] (204.81, 24.50) -- (204.81, 27.40);

\path[draw=drawColor,line width= 0.2pt,line join=round] (238.60, 24.50) -- (238.60, 27.40);

\path[draw=drawColor,line width= 0.2pt,line join=round] (272.39, 24.50) -- (272.39, 27.40);

\path[draw=drawColor,line width= 0.2pt,line join=round] (137.24, 36.07) -- (137.24, 38.96);

\path[draw=drawColor,line width= 0.2pt,line join=round] (171.02, 36.07) -- (171.02, 38.96);

\path[draw=drawColor,line width= 0.2pt,line join=round] (204.81, 36.07) -- (204.81, 38.96);

\path[draw=drawColor,line width= 0.2pt,line join=round] (238.60, 36.07) -- (238.60, 38.96);

\path[draw=drawColor,line width= 0.2pt,line join=round] (272.39, 36.07) -- (272.39, 38.96);
\end{scope}
\end{tikzpicture}
 }
  \end{minipage}
\begin{minipage}[t]{0.3\textwidth}
  \resizebox{1\textwidth}{!}{
    \begin{tikzpicture}[x=1pt,y=1pt]
\definecolor{fillColor}{RGB}{255,255,255}
\path[use as bounding box,fill=fillColor,fill opacity=0.00] (0,0) rectangle (289.08,578.16);
\begin{scope}
\path[clip] ( 55.60, 55.46) rectangle (273.82,535.56);
\definecolor{drawColor}{gray}{0.92}

\path[draw=drawColor,line width= 0.6pt,line join=round] ( 55.60, 77.28) --
	(273.82, 77.28);

\path[draw=drawColor,line width= 0.6pt,line join=round] ( 55.60,120.93) --
	(273.82,120.93);

\path[draw=drawColor,line width= 0.6pt,line join=round] ( 55.60,164.57) --
	(273.82,164.57);

\path[draw=drawColor,line width= 0.6pt,line join=round] ( 55.60,208.22) --
	(273.82,208.22);

\path[draw=drawColor,line width= 0.6pt,line join=round] ( 55.60,251.86) --
	(273.82,251.86);

\path[draw=drawColor,line width= 0.6pt,line join=round] ( 55.60,295.51) --
	(273.82,295.51);

\path[draw=drawColor,line width= 0.6pt,line join=round] ( 55.60,339.15) --
	(273.82,339.15);

\path[draw=drawColor,line width= 0.6pt,line join=round] ( 55.60,382.80) --
	(273.82,382.80);

\path[draw=drawColor,line width= 0.6pt,line join=round] ( 55.60,426.44) --
	(273.82,426.44);

\path[draw=drawColor,line width= 0.6pt,line join=round] ( 55.60,470.09) --
	(273.82,470.09);

\path[draw=drawColor,line width= 0.6pt,line join=round] ( 55.60,513.73) --
	(273.82,513.73);

\path[draw=drawColor,line width= 0.6pt,line join=round] ( 77.42, 55.46) --
	( 77.42,535.56);

\path[draw=drawColor,line width= 0.6pt,line join=round] (121.07, 55.46) --
	(121.07,535.56);

\path[draw=drawColor,line width= 0.6pt,line join=round] (164.71, 55.46) --
	(164.71,535.56);

\path[draw=drawColor,line width= 0.6pt,line join=round] (208.36, 55.46) --
	(208.36,535.56);

\path[draw=drawColor,line width= 0.6pt,line join=round] (252.00, 55.46) --
	(252.00,535.56);
\definecolor{drawColor}{RGB}{255,255,255}
\definecolor{fillColor}{RGB}{255,0,0}

\path[draw=drawColor,line width= 0.1pt,line join=round,fill=fillColor] ( 55.60,491.91) rectangle ( 99.24,535.56);
\definecolor{fillColor}{RGB}{255,54,28}

\path[draw=drawColor,line width= 0.1pt,line join=round,fill=fillColor] ( 55.60,186.39) rectangle ( 99.24,230.04);
\definecolor{fillColor}{RGB}{255,9,2}

\path[draw=drawColor,line width= 0.1pt,line join=round,fill=fillColor] ( 55.60,230.04) rectangle ( 99.24,273.68);
\definecolor{fillColor}{RGB}{255,0,0}

\path[draw=drawColor,line width= 0.1pt,line join=round,fill=fillColor] ( 55.60,404.62) rectangle ( 99.24,448.27);

\path[draw=drawColor,line width= 0.1pt,line join=round,fill=fillColor] ( 55.60,448.27) rectangle ( 99.24,491.91);

\path[draw=drawColor,line width= 0.1pt,line join=round,fill=fillColor] ( 55.60,360.97) rectangle ( 99.24,404.62);
\definecolor{fillColor}{RGB}{255,255,255}

\path[draw=drawColor,line width= 0.1pt,line join=round,fill=fillColor] ( 55.60, 55.46) rectangle ( 99.24, 99.10);
\definecolor{fillColor}{RGB}{255,71,41}

\path[draw=drawColor,line width= 0.1pt,line join=round,fill=fillColor] ( 55.60,142.75) rectangle ( 99.24,186.39);
\definecolor{fillColor}{RGB}{255,6,1}

\path[draw=drawColor,line width= 0.1pt,line join=round,fill=fillColor] ( 55.60,273.68) rectangle ( 99.24,317.33);
\definecolor{fillColor}{RGB}{255,107,73}

\path[draw=drawColor,line width= 0.1pt,line join=round,fill=fillColor] ( 55.60, 99.10) rectangle ( 99.24,142.75);
\definecolor{fillColor}{RGB}{255,0,0}

\path[draw=drawColor,line width= 0.1pt,line join=round,fill=fillColor] ( 55.60,317.33) rectangle ( 99.24,360.97);
\definecolor{fillColor}{gray}{0.50}

\path[draw=drawColor,line width= 0.1pt,line join=round,fill=fillColor] ( 99.24,491.91) rectangle (142.89,535.56);
\definecolor{fillColor}{RGB}{255,0,0}

\path[draw=drawColor,line width= 0.1pt,line join=round,fill=fillColor] ( 99.24,186.39) rectangle (142.89,230.04);
\definecolor{fillColor}{RGB}{255,100,66}

\path[draw=drawColor,line width= 0.1pt,line join=round,fill=fillColor] ( 99.24,230.04) rectangle (142.89,273.68);
\definecolor{fillColor}{RGB}{255,29,11}

\path[draw=drawColor,line width= 0.1pt,line join=round,fill=fillColor] ( 99.24,404.62) rectangle (142.89,448.27);
\definecolor{fillColor}{gray}{0.50}

\path[draw=drawColor,line width= 0.1pt,line join=round,fill=fillColor] ( 99.24,448.27) rectangle (142.89,491.91);

\path[draw=drawColor,line width= 0.1pt,line join=round,fill=fillColor] ( 99.24,360.97) rectangle (142.89,404.62);

\path[draw=drawColor,line width= 0.1pt,line join=round,fill=fillColor] ( 99.24, 55.46) rectangle (142.89, 99.10);
\definecolor{fillColor}{RGB}{255,90,58}

\path[draw=drawColor,line width= 0.1pt,line join=round,fill=fillColor] ( 99.24,142.75) rectangle (142.89,186.39);
\definecolor{fillColor}{RGB}{255,212,197}

\path[draw=drawColor,line width= 0.1pt,line join=round,fill=fillColor] ( 99.24,273.68) rectangle (142.89,317.33);
\definecolor{fillColor}{RGB}{255,242,236}

\path[draw=drawColor,line width= 0.1pt,line join=round,fill=fillColor] ( 99.24, 99.10) rectangle (142.89,142.75);
\definecolor{fillColor}{RGB}{255,0,0}

\path[draw=drawColor,line width= 0.1pt,line join=round,fill=fillColor] ( 99.24,317.33) rectangle (142.89,360.97);
\definecolor{fillColor}{gray}{0.50}

\path[draw=drawColor,line width= 0.1pt,line join=round,fill=fillColor] (142.89,491.91) rectangle (186.53,535.56);
\definecolor{fillColor}{RGB}{255,173,148}

\path[draw=drawColor,line width= 0.1pt,line join=round,fill=fillColor] (142.89,186.39) rectangle (186.53,230.04);
\definecolor{fillColor}{RGB}{255,142,111}

\path[draw=drawColor,line width= 0.1pt,line join=round,fill=fillColor] (142.89,230.04) rectangle (186.53,273.68);
\definecolor{fillColor}{RGB}{255,62,34}

\path[draw=drawColor,line width= 0.1pt,line join=round,fill=fillColor] (142.89,404.62) rectangle (186.53,448.27);
\definecolor{fillColor}{gray}{0.50}

\path[draw=drawColor,line width= 0.1pt,line join=round,fill=fillColor] (142.89,448.27) rectangle (186.53,491.91);
\definecolor{fillColor}{RGB}{255,255,255}

\path[draw=drawColor,line width= 0.1pt,line join=round,fill=fillColor] (142.89,360.97) rectangle (186.53,404.62);

\path[draw=drawColor,line width= 0.1pt,line join=round,fill=fillColor] (142.89, 55.46) rectangle (186.53, 99.10);
\definecolor{fillColor}{RGB}{255,141,110}

\path[draw=drawColor,line width= 0.1pt,line join=round,fill=fillColor] (142.89,142.75) rectangle (186.53,186.39);
\definecolor{fillColor}{RGB}{255,229,219}

\path[draw=drawColor,line width= 0.1pt,line join=round,fill=fillColor] (142.89,273.68) rectangle (186.53,317.33);
\definecolor{fillColor}{RGB}{255,246,242}

\path[draw=drawColor,line width= 0.1pt,line join=round,fill=fillColor] (142.89, 99.10) rectangle (186.53,142.75);
\definecolor{fillColor}{RGB}{255,0,0}

\path[draw=drawColor,line width= 0.1pt,line join=round,fill=fillColor] (142.89,317.33) rectangle (186.53,360.97);
\definecolor{fillColor}{gray}{0.50}

\path[draw=drawColor,line width= 0.1pt,line join=round,fill=fillColor] (186.53,491.91) rectangle (230.18,535.56);
\definecolor{fillColor}{RGB}{255,0,0}

\path[draw=drawColor,line width= 0.1pt,line join=round,fill=fillColor] (186.53,186.39) rectangle (230.18,230.04);
\definecolor{fillColor}{RGB}{255,160,132}

\path[draw=drawColor,line width= 0.1pt,line join=round,fill=fillColor] (186.53,230.04) rectangle (230.18,273.68);
\definecolor{fillColor}{RGB}{255,225,214}

\path[draw=drawColor,line width= 0.1pt,line join=round,fill=fillColor] (186.53,404.62) rectangle (230.18,448.27);
\definecolor{fillColor}{RGB}{255,0,0}

\path[draw=drawColor,line width= 0.1pt,line join=round,fill=fillColor] (186.53,448.27) rectangle (230.18,491.91);
\definecolor{fillColor}{gray}{0.50}

\path[draw=drawColor,line width= 0.1pt,line join=round,fill=fillColor] (186.53,360.97) rectangle (230.18,404.62);

\path[draw=drawColor,line width= 0.1pt,line join=round,fill=fillColor] (186.53, 55.46) rectangle (230.18, 99.10);
\definecolor{fillColor}{RGB}{255,91,58}

\path[draw=drawColor,line width= 0.1pt,line join=round,fill=fillColor] (186.53,142.75) rectangle (230.18,186.39);
\definecolor{fillColor}{RGB}{255,239,233}

\path[draw=drawColor,line width= 0.1pt,line join=round,fill=fillColor] (186.53,273.68) rectangle (230.18,317.33);
\definecolor{fillColor}{RGB}{255,255,254}

\path[draw=drawColor,line width= 0.1pt,line join=round,fill=fillColor] (186.53, 99.10) rectangle (230.18,142.75);
\definecolor{fillColor}{RGB}{255,0,0}

\path[draw=drawColor,line width= 0.1pt,line join=round,fill=fillColor] (186.53,317.33) rectangle (230.18,360.97);
\definecolor{fillColor}{RGB}{255,186,163}

\path[draw=drawColor,line width= 0.1pt,line join=round,fill=fillColor] (230.18,491.91) rectangle (273.82,535.56);
\definecolor{fillColor}{RGB}{255,251,250}

\path[draw=drawColor,line width= 0.1pt,line join=round,fill=fillColor] (230.18,186.39) rectangle (273.82,230.04);
\definecolor{fillColor}{RGB}{255,205,188}

\path[draw=drawColor,line width= 0.1pt,line join=round,fill=fillColor] (230.18,230.04) rectangle (273.82,273.68);
\definecolor{fillColor}{RGB}{255,251,249}

\path[draw=drawColor,line width= 0.1pt,line join=round,fill=fillColor] (230.18,404.62) rectangle (273.82,448.27);
\definecolor{fillColor}{RGB}{255,148,117}

\path[draw=drawColor,line width= 0.1pt,line join=round,fill=fillColor] (230.18,448.27) rectangle (273.82,491.91);
\definecolor{fillColor}{RGB}{255,255,255}

\path[draw=drawColor,line width= 0.1pt,line join=round,fill=fillColor] (230.18,360.97) rectangle (273.82,404.62);

\path[draw=drawColor,line width= 0.1pt,line join=round,fill=fillColor] (230.18, 55.46) rectangle (273.82, 99.10);
\definecolor{fillColor}{RGB}{255,142,111}

\path[draw=drawColor,line width= 0.1pt,line join=round,fill=fillColor] (230.18,142.75) rectangle (273.82,186.39);
\definecolor{fillColor}{RGB}{255,253,252}

\path[draw=drawColor,line width= 0.1pt,line join=round,fill=fillColor] (230.18,273.68) rectangle (273.82,317.33);
\definecolor{fillColor}{RGB}{255,255,255}

\path[draw=drawColor,line width= 0.1pt,line join=round,fill=fillColor] (230.18, 99.10) rectangle (273.82,142.75);
\definecolor{fillColor}{RGB}{255,0,0}

\path[draw=drawColor,line width= 0.1pt,line join=round,fill=fillColor] (230.18,317.33) rectangle (273.82,360.97);
\definecolor{drawColor}{RGB}{0,0,0}

\node[text=drawColor,anchor=base,inner sep=0pt, outer sep=0pt, scale=  2] at ( 77.42,509.93) {1};

\node[text=drawColor,anchor=base,inner sep=0pt, outer sep=0pt, scale=  2] at ( 77.42,204.41) {0.921};

\node[text=drawColor,anchor=base,inner sep=0pt, outer sep=0pt, scale=  2] at ( 77.42,248.06) {0.994};

\node[text=drawColor,anchor=base,inner sep=0pt, outer sep=0pt, scale=  2] at ( 77.42,422.64) {1};

\node[text=drawColor,anchor=base,inner sep=0pt, outer sep=0pt, scale=  2] at ( 77.42,466.29) {1};

\node[text=drawColor,anchor=base,inner sep=0pt, outer sep=0pt, scale=  2] at ( 77.42,379.00) {1};

\node[text=drawColor,anchor=base,inner sep=0pt, outer sep=0pt, scale=  2] at ( 77.42, 73.48) {0};

\node[text=drawColor,anchor=base,inner sep=0pt, outer sep=0pt, scale=  2] at ( 77.42,160.77) {0.871};

\node[text=drawColor,anchor=base,inner sep=0pt, outer sep=0pt, scale=  2] at ( 77.42,291.71) {0.996};

\node[text=drawColor,anchor=base,inner sep=0pt, outer sep=0pt, scale=  2] at ( 77.42,117.12) {0.737};

\node[text=drawColor,anchor=base,inner sep=0pt, outer sep=0pt, scale=  2] at ( 77.42,335.35) {1};

\node[text=drawColor,anchor=base,inner sep=0pt, outer sep=0pt, scale=  2] at (121.07,204.41) {1};

\node[text=drawColor,anchor=base,inner sep=0pt, outer sep=0pt, scale=  2] at (121.07,248.06) {0.767};

\node[text=drawColor,anchor=base,inner sep=0pt, outer sep=0pt, scale=  2] at (121.07,422.64) {0.973};

\node[text=drawColor,anchor=base,inner sep=0pt, outer sep=0pt, scale=  2] at (121.07,160.77) {0.803};

\node[text=drawColor,anchor=base,inner sep=0pt, outer sep=0pt, scale=  2] at (121.07,291.71) {0.224};

\node[text=drawColor,anchor=base,inner sep=0pt, outer sep=0pt, scale=  2] at (121.07,117.12) {0.071};

\node[text=drawColor,anchor=base,inner sep=0pt, outer sep=0pt, scale=  2] at (121.07,335.35) {1};

\node[text=drawColor,anchor=base,inner sep=0pt, outer sep=0pt, scale=  2] at (164.71,204.41) {0.424};

\node[text=drawColor,anchor=base,inner sep=0pt, outer sep=0pt, scale=  2] at (164.71,248.06) {0.577};

\node[text=drawColor,anchor=base,inner sep=0pt, outer sep=0pt, scale=  2] at (164.71,422.64) {0.899};

\node[text=drawColor,anchor=base,inner sep=0pt, outer sep=0pt, scale=  2] at (164.71,379.00) {0};

\node[text=drawColor,anchor=base,inner sep=0pt, outer sep=0pt, scale=  2] at (164.71, 73.48) {0};

\node[text=drawColor,anchor=base,inner sep=0pt, outer sep=0pt, scale=  2] at (164.71,160.77) {0.581};

\node[text=drawColor,anchor=base,inner sep=0pt, outer sep=0pt, scale=  2] at (164.71,291.71) {0.137};

\node[text=drawColor,anchor=base,inner sep=0pt, outer sep=0pt, scale=  2] at (164.71,117.12) {0.048};

\node[text=drawColor,anchor=base,inner sep=0pt, outer sep=0pt, scale=  2] at (164.71,335.35) {1};

\node[text=drawColor,anchor=base,inner sep=0pt, outer sep=0pt, scale=  2] at (208.36,204.41) {1};

\node[text=drawColor,anchor=base,inner sep=0pt, outer sep=0pt, scale=  2] at (208.36,248.06) {0.491};

\node[text=drawColor,anchor=base,inner sep=0pt, outer sep=0pt, scale=  2] at (208.36,422.64) {0.158};

\node[text=drawColor,anchor=base,inner sep=0pt, outer sep=0pt, scale=  2] at (208.36,466.29) {1};

\node[text=drawColor,anchor=base,inner sep=0pt, outer sep=0pt, scale=  2] at (208.36,160.77) {0.8};

\node[text=drawColor,anchor=base,inner sep=0pt, outer sep=0pt, scale=  2] at (208.36,291.71) {0.084};

\node[text=drawColor,anchor=base,inner sep=0pt, outer sep=0pt, scale=  2] at (208.36,117.12) {0.002};

\node[text=drawColor,anchor=base,inner sep=0pt, outer sep=0pt, scale=  2] at (208.36,335.35) {1};

\node[text=drawColor,anchor=base,inner sep=0pt, outer sep=0pt, scale=  2] at (252.00,509.93) {0.36};

\node[text=drawColor,anchor=base,inner sep=0pt, outer sep=0pt, scale=  2] at (252.00,204.41) {0.019};

\node[text=drawColor,anchor=base,inner sep=0pt, outer sep=0pt, scale=  2] at (252.00,248.06) {0.26};

\node[text=drawColor,anchor=base,inner sep=0pt, outer sep=0pt, scale=  2] at (252.00,422.64) {0.021};

\node[text=drawColor,anchor=base,inner sep=0pt, outer sep=0pt, scale=  2] at (252.00,466.29) {0.55};

\node[text=drawColor,anchor=base,inner sep=0pt, outer sep=0pt, scale=  2] at (252.00,379.00) {0};

\node[text=drawColor,anchor=base,inner sep=0pt, outer sep=0pt, scale=  2] at (252.00, 73.48) {0};

\node[text=drawColor,anchor=base,inner sep=0pt, outer sep=0pt, scale=  2] at (252.00,160.77) {0.577};

\node[text=drawColor,anchor=base,inner sep=0pt, outer sep=0pt, scale=  2] at (252.00,291.71) {0.013};

\node[text=drawColor,anchor=base,inner sep=0pt, outer sep=0pt, scale=  2] at (252.00,117.12) {0};

\node[text=drawColor,anchor=base,inner sep=0pt, outer sep=0pt, scale=  2] at (252.00,335.35) {1};
\end{scope}
\begin{scope}
\path[clip] (  0.00,  0.00) rectangle (289.08,578.16);
\definecolor{drawColor}{gray}{0.30}

\node[text=drawColor,anchor=base,inner sep=0pt, outer sep=0pt, scale=  1.5] at ( 77.42,540.51) {A};

\node[text=drawColor,anchor=base,inner sep=0pt, outer sep=0pt, scale=  1.5] at (121.07,540.51) {B};

\node[text=drawColor,anchor=base,inner sep=0pt, outer sep=0pt, scale=  1.5] at (164.71,540.51) {C};

\node[text=drawColor,anchor=base,inner sep=0pt, outer sep=0pt, scale=  1.5] at (208.36,540.51) {D};

\node[text=drawColor,anchor=base,inner sep=0pt, outer sep=0pt, scale=  1.5] at (252.00,540.51) {E};
\end{scope}
\begin{scope}
\path[clip] (  0.00,  0.00) rectangle (289.08,578.16);
\definecolor{drawColor}{gray}{0.30}

\node[text=drawColor,anchor=base east,inner sep=0pt, outer sep=0pt, scale=  1.5] at ( 50.65, 74.25) {11.};

\node[text=drawColor,anchor=base east,inner sep=0pt, outer sep=0pt, scale=  1.5] at ( 50.65,117.90) {10.};

\node[text=drawColor,anchor=base east,inner sep=0pt, outer sep=0pt, scale=  1.5] at ( 50.65,161.54) {9.};

\node[text=drawColor,anchor=base east,inner sep=0pt, outer sep=0pt, scale=  1.5] at ( 50.65,205.19) {8.};

\node[text=drawColor,anchor=base east,inner sep=0pt, outer sep=0pt, scale=  1.5] at ( 50.65,248.83) {7.};

\node[text=drawColor,anchor=base east,inner sep=0pt, outer sep=0pt, scale=  1.5] at ( 50.65,292.48) {6.};

\node[text=drawColor,anchor=base east,inner sep=0pt, outer sep=0pt, scale=  1.5] at ( 50.65,336.12) {5.};

\node[text=drawColor,anchor=base east,inner sep=0pt, outer sep=0pt, scale=  1.5] at ( 50.65,379.77) {4.};

\node[text=drawColor,anchor=base east,inner sep=0pt, outer sep=0pt, scale=  1.5] at ( 50.65,423.41) {3.};

\node[text=drawColor,anchor=base east,inner sep=0pt, outer sep=0pt, scale=  1.5] at ( 50.65,467.06) {2.};

\node[text=drawColor,anchor=base east,inner sep=0pt, outer sep=0pt, scale=  1.5] at ( 50.65,510.70) {1.};
\end{scope}
\begin{scope}
\path[clip] (  0.00,  0.00) rectangle (289.08,578.16);
\definecolor{drawColor}{RGB}{0,0,0}

\node[text=drawColor,anchor=base,inner sep=0pt, outer sep=0pt, scale=  2] at (164.71,563.14) {projection};

\node[text=drawColor,anchor=base,inner sep=0pt, outer sep=0pt, scale=  2] at (164.71,551.26) {};
\end{scope}
\begin{scope}
\path[clip] (  0.00,  0.00) rectangle (289.08,578.16);
\definecolor{drawColor}{RGB}{0,0,0}

\node[text=drawColor,rotate= 90.00,anchor=base,inner sep=0pt, outer sep=0pt, scale=  2] at ( 22.83,295.51) {event log};

\node[text=drawColor,rotate= 90.00,anchor=base,inner sep=0pt, outer sep=0pt, scale=  2] at ( 34.71,295.51) {};
\end{scope}
\begin{scope}
\path[clip] (  0.00,  0.00) rectangle (289.08,578.16);
\definecolor{drawColor}{RGB}{255,255,255}

\path[draw=drawColor,line width= 0.2pt,line join=round] (137.24, 24.50) -- (137.24, 27.40);

\path[draw=drawColor,line width= 0.2pt,line join=round] (171.02, 24.50) -- (171.02, 27.40);

\path[draw=drawColor,line width= 0.2pt,line join=round] (204.81, 24.50) -- (204.81, 27.40);

\path[draw=drawColor,line width= 0.2pt,line join=round] (238.60, 24.50) -- (238.60, 27.40);

\path[draw=drawColor,line width= 0.2pt,line join=round] (272.39, 24.50) -- (272.39, 27.40);

\path[draw=drawColor,line width= 0.2pt,line join=round] (137.24, 36.07) -- (137.24, 38.96);

\path[draw=drawColor,line width= 0.2pt,line join=round] (171.02, 36.07) -- (171.02, 38.96);

\path[draw=drawColor,line width= 0.2pt,line join=round] (204.81, 36.07) -- (204.81, 38.96);

\path[draw=drawColor,line width= 0.2pt,line join=round] (238.60, 36.07) -- (238.60, 38.96);

\path[draw=drawColor,line width= 0.2pt,line join=round] (272.39, 36.07) -- (272.39, 38.96);
\end{scope}
\end{tikzpicture}
 }
  \end{minipage}
  \begin{minipage}[t]{0.3\textwidth}
  \resizebox{1\textwidth}{!}{
    \begin{tikzpicture}[x=1pt,y=1pt]
\definecolor{fillColor}{RGB}{255,255,255}
\path[use as bounding box,fill=fillColor,fill opacity=0.00] (0,0) rectangle (289.08,578.16);
\begin{scope}
\path[clip] ( 55.60, 55.46) rectangle (273.82,535.56);
\definecolor{drawColor}{gray}{0.92}

\path[draw=drawColor,line width= 0.6pt,line join=round] ( 55.60, 77.28) --
	(273.82, 77.28);

\path[draw=drawColor,line width= 0.6pt,line join=round] ( 55.60,120.93) --
	(273.82,120.93);

\path[draw=drawColor,line width= 0.6pt,line join=round] ( 55.60,164.57) --
	(273.82,164.57);

\path[draw=drawColor,line width= 0.6pt,line join=round] ( 55.60,208.22) --
	(273.82,208.22);

\path[draw=drawColor,line width= 0.6pt,line join=round] ( 55.60,251.86) --
	(273.82,251.86);

\path[draw=drawColor,line width= 0.6pt,line join=round] ( 55.60,295.51) --
	(273.82,295.51);

\path[draw=drawColor,line width= 0.6pt,line join=round] ( 55.60,339.15) --
	(273.82,339.15);

\path[draw=drawColor,line width= 0.6pt,line join=round] ( 55.60,382.80) --
	(273.82,382.80);

\path[draw=drawColor,line width= 0.6pt,line join=round] ( 55.60,426.44) --
	(273.82,426.44);

\path[draw=drawColor,line width= 0.6pt,line join=round] ( 55.60,470.09) --
	(273.82,470.09);

\path[draw=drawColor,line width= 0.6pt,line join=round] ( 55.60,513.73) --
	(273.82,513.73);

\path[draw=drawColor,line width= 0.6pt,line join=round] ( 77.42, 55.46) --
	( 77.42,535.56);

\path[draw=drawColor,line width= 0.6pt,line join=round] (121.07, 55.46) --
	(121.07,535.56);

\path[draw=drawColor,line width= 0.6pt,line join=round] (164.71, 55.46) --
	(164.71,535.56);

\path[draw=drawColor,line width= 0.6pt,line join=round] (208.36, 55.46) --
	(208.36,535.56);

\path[draw=drawColor,line width= 0.6pt,line join=round] (252.00, 55.46) --
	(252.00,535.56);
\definecolor{drawColor}{RGB}{255,255,255}
\definecolor{fillColor}{RGB}{255,0,0}

\path[draw=drawColor,line width= 0.1pt,line join=round,fill=fillColor] ( 55.60,491.91) rectangle ( 99.24,535.56);
\definecolor{fillColor}{RGB}{255,54,28}

\path[draw=drawColor,line width= 0.1pt,line join=round,fill=fillColor] ( 55.60,186.39) rectangle ( 99.24,230.04);
\definecolor{fillColor}{RGB}{255,7,2}

\path[draw=drawColor,line width= 0.1pt,line join=round,fill=fillColor] ( 55.60,230.04) rectangle ( 99.24,273.68);
\definecolor{fillColor}{RGB}{255,0,0}

\path[draw=drawColor,line width= 0.1pt,line join=round,fill=fillColor] ( 55.60,404.62) rectangle ( 99.24,448.27);

\path[draw=drawColor,line width= 0.1pt,line join=round,fill=fillColor] ( 55.60,448.27) rectangle ( 99.24,491.91);

\path[draw=drawColor,line width= 0.1pt,line join=round,fill=fillColor] ( 55.60,360.97) rectangle ( 99.24,404.62);
\definecolor{fillColor}{RGB}{255,255,255}

\path[draw=drawColor,line width= 0.1pt,line join=round,fill=fillColor] ( 55.60, 55.46) rectangle ( 99.24, 99.10);
\definecolor{fillColor}{RGB}{255,71,41}

\path[draw=drawColor,line width= 0.1pt,line join=round,fill=fillColor] ( 55.60,142.75) rectangle ( 99.24,186.39);
\definecolor{fillColor}{RGB}{255,6,1}

\path[draw=drawColor,line width= 0.1pt,line join=round,fill=fillColor] ( 55.60,273.68) rectangle ( 99.24,317.33);
\definecolor{fillColor}{RGB}{255,107,73}

\path[draw=drawColor,line width= 0.1pt,line join=round,fill=fillColor] ( 55.60, 99.10) rectangle ( 99.24,142.75);
\definecolor{fillColor}{RGB}{255,0,0}

\path[draw=drawColor,line width= 0.1pt,line join=round,fill=fillColor] ( 55.60,317.33) rectangle ( 99.24,360.97);
\definecolor{fillColor}{gray}{0.50}

\path[draw=drawColor,line width= 0.1pt,line join=round,fill=fillColor] ( 99.24,491.91) rectangle (142.89,535.56);
\definecolor{fillColor}{RGB}{255,0,0}

\path[draw=drawColor,line width= 0.1pt,line join=round,fill=fillColor] ( 99.24,186.39) rectangle (142.89,230.04);
\definecolor{fillColor}{RGB}{255,97,64}

\path[draw=drawColor,line width= 0.1pt,line join=round,fill=fillColor] ( 99.24,230.04) rectangle (142.89,273.68);
\definecolor{fillColor}{RGB}{255,29,11}

\path[draw=drawColor,line width= 0.1pt,line join=round,fill=fillColor] ( 99.24,404.62) rectangle (142.89,448.27);
\definecolor{fillColor}{gray}{0.50}

\path[draw=drawColor,line width= 0.1pt,line join=round,fill=fillColor] ( 99.24,448.27) rectangle (142.89,491.91);

\path[draw=drawColor,line width= 0.1pt,line join=round,fill=fillColor] ( 99.24,360.97) rectangle (142.89,404.62);

\path[draw=drawColor,line width= 0.1pt,line join=round,fill=fillColor] ( 99.24, 55.46) rectangle (142.89, 99.10);
\definecolor{fillColor}{RGB}{255,90,58}

\path[draw=drawColor,line width= 0.1pt,line join=round,fill=fillColor] ( 99.24,142.75) rectangle (142.89,186.39);
\definecolor{fillColor}{RGB}{255,212,197}

\path[draw=drawColor,line width= 0.1pt,line join=round,fill=fillColor] ( 99.24,273.68) rectangle (142.89,317.33);
\definecolor{fillColor}{RGB}{255,242,236}

\path[draw=drawColor,line width= 0.1pt,line join=round,fill=fillColor] ( 99.24, 99.10) rectangle (142.89,142.75);
\definecolor{fillColor}{RGB}{255,0,0}

\path[draw=drawColor,line width= 0.1pt,line join=round,fill=fillColor] ( 99.24,317.33) rectangle (142.89,360.97);
\definecolor{fillColor}{gray}{0.50}

\path[draw=drawColor,line width= 0.1pt,line join=round,fill=fillColor] (142.89,491.91) rectangle (186.53,535.56);
\definecolor{fillColor}{RGB}{255,173,148}

\path[draw=drawColor,line width= 0.1pt,line join=round,fill=fillColor] (142.89,186.39) rectangle (186.53,230.04);
\definecolor{fillColor}{RGB}{255,139,107}

\path[draw=drawColor,line width= 0.1pt,line join=round,fill=fillColor] (142.89,230.04) rectangle (186.53,273.68);
\definecolor{fillColor}{RGB}{255,62,34}

\path[draw=drawColor,line width= 0.1pt,line join=round,fill=fillColor] (142.89,404.62) rectangle (186.53,448.27);
\definecolor{fillColor}{gray}{0.50}

\path[draw=drawColor,line width= 0.1pt,line join=round,fill=fillColor] (142.89,448.27) rectangle (186.53,491.91);
\definecolor{fillColor}{RGB}{255,255,255}

\path[draw=drawColor,line width= 0.1pt,line join=round,fill=fillColor] (142.89,360.97) rectangle (186.53,404.62);

\path[draw=drawColor,line width= 0.1pt,line join=round,fill=fillColor] (142.89, 55.46) rectangle (186.53, 99.10);
\definecolor{fillColor}{RGB}{255,141,110}

\path[draw=drawColor,line width= 0.1pt,line join=round,fill=fillColor] (142.89,142.75) rectangle (186.53,186.39);
\definecolor{fillColor}{RGB}{255,229,219}

\path[draw=drawColor,line width= 0.1pt,line join=round,fill=fillColor] (142.89,273.68) rectangle (186.53,317.33);
\definecolor{fillColor}{RGB}{255,246,242}

\path[draw=drawColor,line width= 0.1pt,line join=round,fill=fillColor] (142.89, 99.10) rectangle (186.53,142.75);
\definecolor{fillColor}{RGB}{255,0,0}

\path[draw=drawColor,line width= 0.1pt,line join=round,fill=fillColor] (142.89,317.33) rectangle (186.53,360.97);
\definecolor{fillColor}{gray}{0.50}

\path[draw=drawColor,line width= 0.1pt,line join=round,fill=fillColor] (186.53,491.91) rectangle (230.18,535.56);
\definecolor{fillColor}{RGB}{255,0,0}

\path[draw=drawColor,line width= 0.1pt,line join=round,fill=fillColor] (186.53,186.39) rectangle (230.18,230.04);
\definecolor{fillColor}{RGB}{255,159,130}

\path[draw=drawColor,line width= 0.1pt,line join=round,fill=fillColor] (186.53,230.04) rectangle (230.18,273.68);
\definecolor{fillColor}{RGB}{255,225,214}

\path[draw=drawColor,line width= 0.1pt,line join=round,fill=fillColor] (186.53,404.62) rectangle (230.18,448.27);
\definecolor{fillColor}{RGB}{255,0,0}

\path[draw=drawColor,line width= 0.1pt,line join=round,fill=fillColor] (186.53,448.27) rectangle (230.18,491.91);
\definecolor{fillColor}{gray}{0.50}

\path[draw=drawColor,line width= 0.1pt,line join=round,fill=fillColor] (186.53,360.97) rectangle (230.18,404.62);

\path[draw=drawColor,line width= 0.1pt,line join=round,fill=fillColor] (186.53, 55.46) rectangle (230.18, 99.10);
\definecolor{fillColor}{RGB}{255,91,58}

\path[draw=drawColor,line width= 0.1pt,line join=round,fill=fillColor] (186.53,142.75) rectangle (230.18,186.39);
\definecolor{fillColor}{RGB}{255,239,233}

\path[draw=drawColor,line width= 0.1pt,line join=round,fill=fillColor] (186.53,273.68) rectangle (230.18,317.33);
\definecolor{fillColor}{RGB}{255,255,254}

\path[draw=drawColor,line width= 0.1pt,line join=round,fill=fillColor] (186.53, 99.10) rectangle (230.18,142.75);
\definecolor{fillColor}{RGB}{255,0,0}

\path[draw=drawColor,line width= 0.1pt,line join=round,fill=fillColor] (186.53,317.33) rectangle (230.18,360.97);
\definecolor{fillColor}{RGB}{255,170,144}

\path[draw=drawColor,line width= 0.1pt,line join=round,fill=fillColor] (230.18,491.91) rectangle (273.82,535.56);
\definecolor{fillColor}{RGB}{255,251,249}

\path[draw=drawColor,line width= 0.1pt,line join=round,fill=fillColor] (230.18,186.39) rectangle (273.82,230.04);
\definecolor{fillColor}{RGB}{255,204,187}

\path[draw=drawColor,line width= 0.1pt,line join=round,fill=fillColor] (230.18,230.04) rectangle (273.82,273.68);
\definecolor{fillColor}{RGB}{255,251,249}

\path[draw=drawColor,line width= 0.1pt,line join=round,fill=fillColor] (230.18,404.62) rectangle (273.82,448.27);
\definecolor{fillColor}{RGB}{255,127,94}

\path[draw=drawColor,line width= 0.1pt,line join=round,fill=fillColor] (230.18,448.27) rectangle (273.82,491.91);
\definecolor{fillColor}{RGB}{255,255,255}

\path[draw=drawColor,line width= 0.1pt,line join=round,fill=fillColor] (230.18,360.97) rectangle (273.82,404.62);

\path[draw=drawColor,line width= 0.1pt,line join=round,fill=fillColor] (230.18, 55.46) rectangle (273.82, 99.10);
\definecolor{fillColor}{RGB}{255,142,111}

\path[draw=drawColor,line width= 0.1pt,line join=round,fill=fillColor] (230.18,142.75) rectangle (273.82,186.39);
\definecolor{fillColor}{RGB}{255,252,251}

\path[draw=drawColor,line width= 0.1pt,line join=round,fill=fillColor] (230.18,273.68) rectangle (273.82,317.33);
\definecolor{fillColor}{RGB}{255,255,255}

\path[draw=drawColor,line width= 0.1pt,line join=round,fill=fillColor] (230.18, 99.10) rectangle (273.82,142.75);
\definecolor{fillColor}{RGB}{255,0,0}

\path[draw=drawColor,line width= 0.1pt,line join=round,fill=fillColor] (230.18,317.33) rectangle (273.82,360.97);
\definecolor{drawColor}{RGB}{0,0,0}

\node[text=drawColor,anchor=base,inner sep=0pt, outer sep=0pt, scale=  2] at ( 77.42,509.93) {1};

\node[text=drawColor,anchor=base,inner sep=0pt, outer sep=0pt, scale=  2] at ( 77.42,204.41) {0.922};

\node[text=drawColor,anchor=base,inner sep=0pt, outer sep=0pt, scale=  2] at ( 77.42,248.06) {0.995};

\node[text=drawColor,anchor=base,inner sep=0pt, outer sep=0pt, scale=  2] at ( 77.42,422.64) {1};

\node[text=drawColor,anchor=base,inner sep=0pt, outer sep=0pt, scale=  2] at ( 77.42,466.29) {1};

\node[text=drawColor,anchor=base,inner sep=0pt, outer sep=0pt, scale=  2] at ( 77.42,379.00) {1};

\node[text=drawColor,anchor=base,inner sep=0pt, outer sep=0pt, scale=  2] at ( 77.42, 73.48) {0};

\node[text=drawColor,anchor=base,inner sep=0pt, outer sep=0pt, scale=  2] at ( 77.42,160.77) {0.871};

\node[text=drawColor,anchor=base,inner sep=0pt, outer sep=0pt, scale=  2] at ( 77.42,291.71) {0.996};

\node[text=drawColor,anchor=base,inner sep=0pt, outer sep=0pt, scale=  2] at ( 77.42,117.12) {0.737};

\node[text=drawColor,anchor=base,inner sep=0pt, outer sep=0pt, scale=  2] at ( 77.42,335.35) {1};

\node[text=drawColor,anchor=base,inner sep=0pt, outer sep=0pt, scale=  2] at (121.07,204.41) {1};

\node[text=drawColor,anchor=base,inner sep=0pt, outer sep=0pt, scale=  2] at (121.07,248.06) {0.778};

\node[text=drawColor,anchor=base,inner sep=0pt, outer sep=0pt, scale=  2] at (121.07,422.64) {0.973};

\node[text=drawColor,anchor=base,inner sep=0pt, outer sep=0pt, scale=  2] at (121.07,160.77) {0.804};

\node[text=drawColor,anchor=base,inner sep=0pt, outer sep=0pt, scale=  2] at (121.07,291.71) {0.224};

\node[text=drawColor,anchor=base,inner sep=0pt, outer sep=0pt, scale=  2] at (121.07,117.12) {0.071};

\node[text=drawColor,anchor=base,inner sep=0pt, outer sep=0pt, scale=  2] at (121.07,335.35) {1};

\node[text=drawColor,anchor=base,inner sep=0pt, outer sep=0pt, scale=  2] at (164.71,204.41) {0.424};

\node[text=drawColor,anchor=base,inner sep=0pt, outer sep=0pt, scale=  2] at (164.71,248.06) {0.592};

\node[text=drawColor,anchor=base,inner sep=0pt, outer sep=0pt, scale=  2] at (164.71,422.64) {0.9};

\node[text=drawColor,anchor=base,inner sep=0pt, outer sep=0pt, scale=  2] at (164.71,379.00) {0};

\node[text=drawColor,anchor=base,inner sep=0pt, outer sep=0pt, scale=  2] at (164.71, 73.48) {0};

\node[text=drawColor,anchor=base,inner sep=0pt, outer sep=0pt, scale=  2] at (164.71,160.77) {0.581};

\node[text=drawColor,anchor=base,inner sep=0pt, outer sep=0pt, scale=  2] at (164.71,291.71) {0.137};

\node[text=drawColor,anchor=base,inner sep=0pt, outer sep=0pt, scale=  2] at (164.71,117.12) {0.048};

\node[text=drawColor,anchor=base,inner sep=0pt, outer sep=0pt, scale=  2] at (164.71,335.35) {1};

\node[text=drawColor,anchor=base,inner sep=0pt, outer sep=0pt, scale=  2] at (208.36,204.41) {1};

\node[text=drawColor,anchor=base,inner sep=0pt, outer sep=0pt, scale=  2] at (208.36,248.06) {0.496};

\node[text=drawColor,anchor=base,inner sep=0pt, outer sep=0pt, scale=  2] at (208.36,422.64) {0.159};

\node[text=drawColor,anchor=base,inner sep=0pt, outer sep=0pt, scale=  2] at (208.36,466.29) {1};

\node[text=drawColor,anchor=base,inner sep=0pt, outer sep=0pt, scale=  2] at (208.36,160.77) {0.8};

\node[text=drawColor,anchor=base,inner sep=0pt, outer sep=0pt, scale=  2] at (208.36,291.71) {0.084};

\node[text=drawColor,anchor=base,inner sep=0pt, outer sep=0pt, scale=  2] at (208.36,117.12) {0.002};

\node[text=drawColor,anchor=base,inner sep=0pt, outer sep=0pt, scale=  2] at (208.36,335.35) {1};

\node[text=drawColor,anchor=base,inner sep=0pt, outer sep=0pt, scale=  2] at (252.00,509.93) {0.44};

\node[text=drawColor,anchor=base,inner sep=0pt, outer sep=0pt, scale=  2] at (252.00,204.41) {0.022};

\node[text=drawColor,anchor=base,inner sep=0pt, outer sep=0pt, scale=  2] at (252.00,248.06) {0.265};

\node[text=drawColor,anchor=base,inner sep=0pt, outer sep=0pt, scale=  2] at (252.00,422.64) {0.023};

\node[text=drawColor,anchor=base,inner sep=0pt, outer sep=0pt, scale=  2] at (252.00,466.29) {0.65};

\node[text=drawColor,anchor=base,inner sep=0pt, outer sep=0pt, scale=  2] at (252.00,379.00) {0};

\node[text=drawColor,anchor=base,inner sep=0pt, outer sep=0pt, scale=  2] at (252.00, 73.48) {0};

\node[text=drawColor,anchor=base,inner sep=0pt, outer sep=0pt, scale=  2] at (252.00,160.77) {0.577};

\node[text=drawColor,anchor=base,inner sep=0pt, outer sep=0pt, scale=  2] at (252.00,291.71) {0.015};

\node[text=drawColor,anchor=base,inner sep=0pt, outer sep=0pt, scale=  2] at (252.00,117.12) {0};

\node[text=drawColor,anchor=base,inner sep=0pt, outer sep=0pt, scale=  2] at (252.00,335.35) {1};
\end{scope}
\begin{scope}
\path[clip] (  0.00,  0.00) rectangle (289.08,578.16);
\definecolor{drawColor}{gray}{0.30}

\node[text=drawColor,anchor=base,inner sep=0pt, outer sep=0pt, scale=  1.5] at ( 77.42,540.51) {A};

\node[text=drawColor,anchor=base,inner sep=0pt, outer sep=0pt, scale=  1.5] at (121.07,540.51) {B};

\node[text=drawColor,anchor=base,inner sep=0pt, outer sep=0pt, scale=  1.5] at (164.71,540.51) {C};

\node[text=drawColor,anchor=base,inner sep=0pt, outer sep=0pt, scale=  1.5] at (208.36,540.51) {D};

\node[text=drawColor,anchor=base,inner sep=0pt, outer sep=0pt, scale=  1.5] at (252.00,540.51) {E};
\end{scope}
\begin{scope}
\path[clip] (  0.00,  0.00) rectangle (289.08,578.16);
\definecolor{drawColor}{gray}{0.30}

\node[text=drawColor,anchor=base east,inner sep=0pt, outer sep=0pt, scale=  1.5] at ( 50.65, 74.25) {11.};

\node[text=drawColor,anchor=base east,inner sep=0pt, outer sep=0pt, scale=  1.5] at ( 50.65,117.90) {10.};

\node[text=drawColor,anchor=base east,inner sep=0pt, outer sep=0pt, scale=  1.5] at ( 50.65,161.54) {9.};

\node[text=drawColor,anchor=base east,inner sep=0pt, outer sep=0pt, scale=  1.5] at ( 50.65,205.19) {8.};

\node[text=drawColor,anchor=base east,inner sep=0pt, outer sep=0pt, scale=  1.5] at ( 50.65,248.83) {7.};

\node[text=drawColor,anchor=base east,inner sep=0pt, outer sep=0pt, scale=  1.5] at ( 50.65,292.48) {6.};

\node[text=drawColor,anchor=base east,inner sep=0pt, outer sep=0pt, scale=  1.5] at ( 50.65,336.12) {5.};

\node[text=drawColor,anchor=base east,inner sep=0pt, outer sep=0pt, scale=  1.5] at ( 50.65,379.77) {4.};

\node[text=drawColor,anchor=base east,inner sep=0pt, outer sep=0pt, scale=  1.5] at ( 50.65,423.41) {3.};

\node[text=drawColor,anchor=base east,inner sep=0pt, outer sep=0pt, scale=  1.5] at ( 50.65,467.06) {2.};

\node[text=drawColor,anchor=base east,inner sep=0pt, outer sep=0pt, scale=  1.5] at ( 50.65,510.70) {1.};
\end{scope}
\begin{scope}
\path[clip] (  0.00,  0.00) rectangle (289.08,578.16);
\definecolor{drawColor}{RGB}{0,0,0}

\node[text=drawColor,anchor=base,inner sep=0pt, outer sep=0pt, scale=  2] at (164.71,563.14) {projection};

\node[text=drawColor,anchor=base,inner sep=0pt, outer sep=0pt, scale=  2] at (164.71,551.26) {};
\end{scope}
\begin{scope}
\path[clip] (  0.00,  0.00) rectangle (289.08,578.16);
\definecolor{drawColor}{RGB}{0,0,0}

\node[text=drawColor,rotate= 90.00,anchor=base,inner sep=0pt, outer sep=0pt, scale=  2] at ( 22.83,295.51) {event log};

\node[text=drawColor,rotate= 90.00,anchor=base,inner sep=0pt, outer sep=0pt, scale=  2] at ( 34.71,295.51) {};
\end{scope}
\begin{scope}
\path[clip] (  0.00,  0.00) rectangle (289.08,578.16);
\definecolor{drawColor}{RGB}{255,255,255}

\path[draw=drawColor,line width= 0.2pt,line join=round] (137.24, 24.50) -- (137.24, 27.40);

\path[draw=drawColor,line width= 0.2pt,line join=round] (171.02, 24.50) -- (171.02, 27.40);

\path[draw=drawColor,line width= 0.2pt,line join=round] (204.81, 24.50) -- (204.81, 27.40);

\path[draw=drawColor,line width= 0.2pt,line join=round] (238.60, 24.50) -- (238.60, 27.40);

\path[draw=drawColor,line width= 0.2pt,line join=round] (272.39, 24.50) -- (272.39, 27.40);

\path[draw=drawColor,line width= 0.2pt,line join=round] (137.24, 36.07) -- (137.24, 38.96);

\path[draw=drawColor,line width= 0.2pt,line join=round] (171.02, 36.07) -- (171.02, 38.96);

\path[draw=drawColor,line width= 0.2pt,line join=round] (204.81, 36.07) -- (204.81, 38.96);

\path[draw=drawColor,line width= 0.2pt,line join=round] (238.60, 36.07) -- (238.60, 38.96);

\path[draw=drawColor,line width= 0.2pt,line join=round] (272.39, 36.07) -- (272.39, 38.96);
\end{scope}
\end{tikzpicture}
 }
  \end{minipage}
\end{minipage}
\begin{minipage}[t]{\textwidth}
\resizebox{1\textwidth}{!}{
\centering
\begin{tikzpicture}[x=1pt,y=1pt]
\definecolor{fillColor}{RGB}{255,255,255}
\path[use as bounding box,fill=fillColor,fill opacity=0.00] (0,0) rectangle (300,30);
\begin{scope}
\path[clip] (  0.00,  0.00) rectangle (300,30);
\node[inner sep=0pt,outer sep=0pt,anchor=south west,rotate=  0.00] at (103.68,  26.81) {
	\pgfimage[width=142.26pt,height= 14.45pt,interpolate=true]{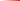}};
\end{scope}
\begin{scope}
\path[clip] (  0.00,  0.00) rectangle (300,30);
\definecolor{drawColor}{RGB}{0,0,0}

\node[text=drawColor,anchor=base,inner sep=0pt, outer sep=0pt, scale=  0.88] at (107.24, 15.25) {0.00};

\node[text=drawColor,anchor=base,inner sep=0pt, outer sep=0pt, scale=  0.88] at (141.02, 15.25) {0.25};

\node[text=drawColor,anchor=base,inner sep=0pt, outer sep=0pt, scale=  0.88] at (174.81, 15.25) {0.50};

\node[text=drawColor,anchor=base,inner sep=0pt, outer sep=0pt, scale=  0.88] at (208.60, 15.25) {0.75};

\node[text=drawColor,anchor=base,inner sep=0pt, outer sep=0pt, scale=  0.88] at (242.39, 15.25) {1.00};
\end{scope}
\begin{scope}
\path[clip] (  0.00,  0.00) rectangle (300,30);
\definecolor{drawColor}{RGB}{0,0,0}

\node[text=drawColor,anchor=base west,inner sep=0pt, outer sep=0pt, scale=  1] at ( 49.48, 24.50) {uniqueness};
\end{scope}
\begin{scope}
\path[clip] (  0.00,  0.00) rectangle (300,30);
\definecolor{drawColor}{RGB}{255,255,255}

\path[draw=drawColor,line width= 0.2pt,line join=round] (107.24, 26.81) -- (107.24, 29.71);

\path[draw=drawColor,line width= 0.2pt,line join=round] (141.02, 26.81) -- (141.02, 29.71);

\path[draw=drawColor,line width= 0.2pt,line join=round] (174.81, 26.81) -- (174.81, 29.71);

\path[draw=drawColor,line width= 0.2pt,line join=round] (208.60, 26.81) -- (208.60, 29.71);

\path[draw=drawColor,line width= 0.2pt,line join=round] (242.39, 26.81) -- (242.39, 29.71);

\path[draw=drawColor,line width= 0.2pt,line join=round] (107.24, 38.38) -- (107.24, 41.27);

\path[draw=drawColor,line width= 0.2pt,line join=round] (141.02, 38.38) -- (141.02, 41.27);

\path[draw=drawColor,line width= 0.2pt,line join=round] (174.81, 38.38) -- (174.81, 41.27);

\path[draw=drawColor,line width= 0.2pt,line join=round] (208.60, 38.38) -- (208.60, 41.27);

\path[draw=drawColor,line width= 0.2pt,line join=round] (242.39, 38.38) -- (242.39, 41.27);
\end{scope}

\end{tikzpicture}
 }
\end{minipage}
\begin{minipage}[t]{\textwidth}
\captionsetup[sub]{labelformat=parens}
\begingroup
    \captionsetup{type=figure}
    \begin{subfigure}{0.3\textwidth}
        \vspace*{-0.5cm}
        \caption{For 10\% of points.}
        \label{fig:uniqueness_all_10}
    \end{subfigure}
    \begin{subfigure}{0.3\textwidth}
        \vspace*{-0.5cm}
        \caption{For 50\% of points.}
        \label{fig:uniqueness_all_50}
    \end{subfigure}
    \begin{subfigure}{0.3\textwidth}
        \vspace*{-0.5cm}
        \caption{For 90\% of points.}
        \label{fig:uniqueness_all_90}
    \end{subfigure}
\endgroup
\end{minipage}
\begin{minipage}[t]{\textwidth}
  \caption{Uniqueness based on traces for all event logs.}
  \label{fig:uniqueness_all}
\end{minipage}
\vspace*{-2em}
\end{figure}

Our method of estimating the uniqueness based on traces can be applied
to all event logs categorized as real-life-individuals (R).
\Cref{fig:uniqueness_all} presents the uniqueness for all event logs
for different projections.
We evaluate the uniqueness given 10\%, 50\%, and 90\% of possible points per trace,
i.e., an adversary knows this number of points per case.
Grey fields without numbers imply that this projection could not be
evaluated due to missing attributes. 

In \Cref{fig:uniqueness_all} we observe a similar trend as before
for the Sepsis event log:
Projection~A generally leads to a high uniqueness.
By omitting information, expressed by the various projections, the uniqueness decreases.
This becomes apparent when comparing projection~B to C,
where the case attributes are removed.
Projection~E, i.e., considering the activities only, leads to a small uniqueness,
with the exception of event log~5 and 9.
We explain this by the fact that these event logs have many different activities
and have a varying trace length per case.
For event log~10, we can already see a clear reduction of the uniqueness
for projection~B. This can be explained by the small number of case attributes
and small number of unique activities.

The most surprising event log is~11. It has no unique cases.
The prime reason for this difference is the result of a timestamp
in daily resolution and the small number of unique activities. It is worth adding that increasing the number of points from 10\% to 50\% is significant with respect to uniqueness compared to the number of points from 50\% to 90\%.
For example, the uniqueness of projection~A for event log~10 increases
from 62.4\% in \Cref{fig:uniqueness_all_10} to 73.7\% in \Cref{fig:uniqueness_all_50}.
Given 90\% of points of the trace, we cannot observe an increase of the uniqueness for event log~10.
This can also be observed for other event logs and other projections.
The prime cause of this is the high variance of the trace length.

Overall in our study, we find that the uniqueness based on traces is higher than on case attributes
(cf. results in \Cref{tab:results_qi}).
For example, event log~3 has a sample uniqueness based on case attributes of 1.1\%.
Based on traces, however, it reaches for projection~C a case uniqueness of 84.4\%.
We conclude that traces are particularly vulnerable to data re-identification attacks.

\subsection{Discussion}
\label{discuss}

Our results demonstrate that 11 of 12 evaluated event logs have a uniqueness greater than 62\%,
even for a random selection of trace points.
More specific information, e.g., the order of individual activities, can lead to a greater uniqueness with fewer points.
Additional knowledge about the process in general could be used by an adversary to predict certain activities,
which was also confirmed in~\cite{pikaPrivacyPreservingProcessMining2019}.
The random selection, however, clearly shows that little background knowledge is sufficient
and already induces a considerable re-identification risk for event logs. In contrast, generalization of attributes helps to reduce the risk~\cite{zook2017ten}.
The results, however, show that combining several attributes, such as case attributes and activities,
still yields unique cases.
In combination with lowering the resolution of values,
e.g., publishing only the year of birth instead of the full birthday,
reduces the re-identification risk.
Such generalization techniques can also be applied to timestamps, activities, or case attributes.
Along the lines of the data minimization principle,
i.e., limiting the amount of personal data,
omitting data is simply the most profound way to reduce the risk,
which we clearly see when taking our projections into account.
Consequently, the projections can be used to reduce the re-identification risk.

We apply our methods to already published event logs to point out the risk of
re-identification in the domain of process mining.
To this end, we only quantify the risk and refrain from cross-correlating other event logs,
which might re-identify individuals.
In addition, we take measures such as pseudonymizing event logs in our evaluation
to neither expose nor blame specific event logs.
 \section{Related Work}
\label{related}
\paragraph{Re-Identification Attacks}
Re-identification attacks were addressed and successfully carried out in the past by a large
number of researchers~\cite{narayanan08netflix,Rocher2019,song2014not,de2013unique,7796899,narayanan2009anonymizing}.
Narayanan and Shmatikov~\cite{narayanan08netflix} de-anonymize a data
set from Netflix containing movie ratings by cross-correlating multiple data sets.
In~\cite{narayanan2009anonymizing}, they modified their approach to apply it to social networks.
In contrast, our adversary's goal is to re-identify an individual (also known as singling out)
and not reconstruct all attribute values of an individual. We therefore measure the uniqueness. We base our uniqueness measures on two well-known approaches~\cite{Rocher2019,de2013unique,song2014not}
and adapt them for the domain of event logs and process mining.
Rocher et al.~\cite{Rocher2019} estimate the population uniqueness based on given attribute
values. We employ their method to estimate the uniqueness based on case attributes.
Our method to estimate the uniqueness based on traces relies on the approach presented in~\cite{de2013unique,song2014not}, where uniqueness in mobility traces with location data is estimated.
Due to the structure of an event log, both methods alone are not sufficient to
determine the uniqueness in event logs and require data preparation.
For example, event logs have a specific format that requires transformation
in order to apply uniqueness measures on traces.

\textit{Privacy in process mining} Awareness of privacy issues in process mining has increased~\cite{spiekermannEngineeringPrivacy2009}, particularly since the General Data Protection Regulation (GDPR) was put into effect. Although the Process Mining Manifesto~\cite{vanderaalstProcessMiningManifesto2012} demands to balance utility and privacy in process mining applications, the number of related contributions is still rather small.
To preserve privacy in event logs while still discovering the correct main 
process behavior has been addressed by Fahrenkrog-Petersen et 
al.~\cite{icpm/Fahrenkrog-Petersen19}. Their algorithm guarantees $k$-anonymity 
and $t$-closeness while maximizing the utility of the sanitized event log. In 
general, $k$-anonymity aggregates the data in such a way that each individual 
cannot be distinguished based on its values from at least $k-1$ other 
individuals of the data 
set~\cite{sweeney2002k,samaratiProtectingPrivacyWhen1998}. Yet, it has been 
shown in the past that neither $k$-anonymity, nor $t$-closeness are sufficient 
to provide strong privacy guarantees~\cite{dwork2006calibrating}.

The strongest privacy model available to date, which provides provable privacy 
guarantees, is differential privacy. It was recently incorporated in a first 
privacy-preserving technique for process mining~\cite{Mannhardt2019}. The 
approach presents a privacy engine capable of keeping personal data private by 
adding 
noise to queries. The privacy techniques 
of~\cite{icpm/Fahrenkrog-Petersen19,Mannhardt2019} have been combined in a 
web-based tool~\cite{bauerELPaaSEventLog2019}. Pseudonymization of data sets 
related to process mining has been discussed 
in~\cite{rafieiEnsuringConfidentialityProcess,burattinAnonymousProcessMining2015}.
 Values of the original data set is replaced with pseudonyms. However, the 
encryption still allows for a potential re-identification by an adversary with 
knowledge about the domain and the statistical distribution of the encrypted 
data.
Beside technological privacy challenges for process mining, the approach of~\cite{mannhardtPrivacyChallengesProcess} also discuss organizational privacy challenges by means of a framework. Although, the approach points to several privacy concerns in process mining, no technical solution is presented. Pika et al.~\cite{pikaPrivacyPreservingProcessMining2019} assess the suitability of existing privacy-preserving approaches for process mining data. They propose a framework to support privacy-preserving process mining analysis. While Pika et al. analyze the suitability of existing data transformation approaches to anonymize process data, they do not provide an approach to support the identification of information, e.g., atypical process behavior, that should be suppressed to reduce the re-identification risk of subjects. Our metric fills this gap and helps data owners to identify the unique cases with atypical process behavior.

In comparison to existing related works on privacy-aware approaches for process mining,
this paper makes an attempt to quantify the re-identification risk. Data publishers can
determine which information should be suppressed before releasing an event log
for process mining. If a high re-identification risk is detected, the approaches mentioned above
might be able to lower the risk of re-identification and therefore to provide higher privacy guarantees.

\section{Conclusion}
\label{conclude}
This paper identifies and evaluates the risk of re-identification in event logs for process mining.
We reveal that there is a serious privacy leakage in the vast majority
of the event logs used widely in the community.
To address this issue, we argue for the use of methods to estimate the uniqueness that allow event log publishers to carefully evaluate their event logs before release and if need to
suppress certain information. Overall, real-world data traces are an essential means to evaluate and compare algorithms. This paper shows that we as a community have to act more carefully, though, when releasing event logs, while also highlighting the need to develop privacy-preserving techniques for event logs.
We believe that this work will foster the trust and increases the willingness for sharing event logs while providing privacy guarantees.

\subsubsection*{Acknowledgement}
We thank Yves-Alexandre de Montjoye for providing his code to estimate data 
uniqueness, which we used in our evaluation.

\bibliographystyle{splncs}

\end{document}